\documentclass[twocolumn,preprintnumbers,amsmath,amssymb,amsfonts]{revtex4-1}
\usepackage{epsfig}
\usepackage{graphicx}

\begin{document}
\preprint{UTTG-08-11}
\preprint{TCC-012-11}

\title{Adiabaticity and the Fate of Non-Gaussianities: The Trispectrum and Beyond}

\author{Joel Meyers}
\email{joelmeyers@mail.utexas.edu}
\author{Navin Sivanandam}
\email{navin.sivanandam@gmail.com}
\affiliation{Theory Group, Department of Physics, University of Texas, Austin, TX 78712}
\affiliation{Texas Cosmology Center, University of Texas, Austin, TX 78712}

\date{\today}

\begin{abstract}
Extending the analysis of \cite{Meyers:2010rg} beyond the bispectrum, we explore the superhorizon generation of local non-gaussianities and their subsequent approach to adiabaticity. Working with a class of two field models of inflation with potentials amenable to treatment with the $\delta N$ formalism we find that, as is the case for $f_{\mathrm{NL}}^{\textrm{local}}$, the local trispectrum parameters $\tau_{\mathrm{NL}}$ and $g_{\mathrm{NL}}$ are exponentially driven toward values which are slow roll suppressed if the fluctuations are driven into an adiabatic mode by a phase of effectively single field inflation. We argue that general considerations should ensure that a similar behavior will hold for the local forms of higher point correlations as well.
\end{abstract}

\maketitle

\section{Introduction}
The observed power spectrum of temperature fluctuations in the cosmic microwave background radiation \cite{Komatsu:2010fb} has provided dramatic evidence for inflation as the origin of cosmological fluctuations \cite{Mukhanov:1981xt,Hawking:1982cz,Starobinsky:1982ee,Guth:1982ec,Bardeen:1983qw,Fischler:1985ky}. However, probing the inflationary paradigm in detail requires that we study more than just the two-point correlation function; we must also examine the non-gaussian contributions to the primordial fluctuations.

If we model inflation as being driven by canonically normalized slowly rolling scalar fields, these fields will generally have gaussian fluctuations at horizon exit \cite{Seery:2005gb,Gao:2008dt}. For single field models, local form non-gaussian contributions to the curvature perturbations are slow-roll suppressed \cite{Maldacena:2002vr,Seery:2005wm}. With multiple fields this may not be the case. The continued evolution of the curvature perturbation outside the horizon can mix initially gaussian scalar fluctuations to produce a large non-gaussian component in the spectrum. In particular, it has been argued that the quantity $f_{\mathrm{NL}}^{\textrm{local}}$ (characterizing the local form bispectrum) can become large in a number of multifield models, through the evolution of the curvature perturbation after horizon exit; for a non-exhaustive list of examples see \cite{Linde:1996gt,Bernardeau:2002jy,Bernardeau:2002jf,Lyth:2002my,Byrnes:2009qy,Battefeld:2006sz,Chen:2009we,Sasaki:2008uc,Battefeld:2009ym,Byrnes:2008zy,Cai:2009hw,Dvali:2003em,Kofman:2003nx,Rigopoulos:2005ae,Rigopoulos:2005us}, or for reviews see \cite{Byrnes:2010em,Wands:2010af}.

Since the curvature perturbation continues to evolve outside the horizon in multifield inflation, to make predictions we should follow inflationary observables until they become conserved -- a process that occurs once cosmological fluctuations enter a purely adiabatic mode \cite{Weinberg:2008zzc,Weinberg:2003sw,Weinberg:2004kr,Weinberg:2008nf} -- or until they are observed.  It is important to stress the difference between conservation of the curvature perturbation, and its intermittent constancy during some periods of the early universe. By conservation we mean that for a particular mode of wave number $k$ the curvature perturbation remains constant while the mode is outside the horizon $\frac{k}{a}\ll H$. The curvature perturbation may become constant without being conserved, as is the case when multiple field inflation takes place along a straight trajectory in field space, or when there exist two decoupled fluids with the same equation of state. The important distinction in these cases is that the curvature perturbation may evolve during a later period if non-adiabatic fluctuations are present when the inflaton traverses a turn, or when one decoupled fluid changes its equation of state, say by becoming non-relativistic. We will focus in this paper on the conservation of the curvature perturbation, as this guarantees that correlations are insensitive to the subsequent evolution of the universe. In a recent paper \cite{Meyers:2010rg} we explored the effect of the approach to adiabaticity on non-gaussianities produced by superhorizon evolution in multifield inflation. In the limited, but illustrative, class of models where isocurvature is eliminated by a short period of single field inflation before reheating, we found that the approach to adiabaticity damped out the leading contribution to $f_{\mathrm{NL}}^{\textrm{local}}$. Our goal in this paper is to carry the analysis further and consider the evolution of higher order correlation functions in the same class of models.

In addition to allowing for superhorizon evolution of the curvature perturbation, non-adiabatic fluctuations which persist through the radiation-dominated era leave observable effects on the cosmic microwave background \cite{Bucher:1999re,Bucher:2000kb,Bucher:2000cd,Bucher:2000hy}.  Current observations are consistent with a purely adiabatic power spectrum, though they allow for a small contribution from non-adibatic fluctuations \cite{Komatsu:2010fb}.

Although higher order correlations come with more significant observational challenges, there are weak bounds on some limits of the trispectrum, bounds which should improve considerably with future experiments. From the 5 year WMAP data Smidt et al.~ \cite{Smidt:2010sv} have found: $-7.4<g_{\mathrm{NL}}/10^5<8.2$ and $-0.6<\tau_{\mathrm{NL}}/10^4<3.3$, and Fergusson, Regan, and Shellard \cite{Fergusson:2010gn} found $-5.4<g_{\mathrm{NL}}/10^5<8.6$. $g_{\mathrm{NL}}$ and $\tau_{\mathrm{NL}}$ characterize the trispectrum in a particular ``local-form'' parametrization \cite{Byrnes:2006vq,Komatsu:2010hc}, the details of which we will discuss below. Upcoming experiments will do significantly better -- the Planck satellite should be able to achieve $\Delta\tau_{\mathrm{NL}}\sim560$ \cite{Komatsu:2010hc}, and probes of large scale structure may be able to provide even tighter bounds \cite{Jeong:2008rj,Giannantonio:2009ak,Sefusatti:2009qh}.

One might, with some apparent justification, wonder why we should bother with the trispectrum. After all the bispectrum is guaranteed to vanish in the presence of purely gaussian fluctuations making it a sensitive probe of non-gaussianity and, given the challenges in measuring higher order correlations, one might believe it unlikely that we would gain much by the effort. This is not, however, the case. Because of the existence of consistency relations between the bispectrum and trispectrum, measurement of $\tau_{\mathrm{NL}}$ can be a sharper probe of non-gaussianity than $f_{\mathrm{NL}}$ in some models (if $f_{\mathrm{NL}}\gtrsim50$) and even holds the possibility of ruling out large classes of multifield models. For this reason, we would like to explore the evolution of the trispectrum in multifield inflationary models as adiabaticity is approached.

As in our previous work we use the $\delta N$ formalism to follow the evolution of perturbations outside the horizon until the universe passes through a phase of single field inflation that drives the isocurvature to zero. We should also stress at this juncture that a phase of single field inflation is not the only way to force perturbations into an adiabatic mode. In models where the non-adiabatic fluctuations persist through the end of inflation, the fluctuations may become adiabatic during a phase of local thermal equilibrium with no non-zero conserved quantum numbers \cite{Weinberg:2004kf,Weinberg:2008si}. There are also other ways to achieve adiabaticity, such as the curvaton scenario \cite{Linde:1996gt,Lyth:2002my}. It is reasonable to suggest that in these models non-gaussianities generated from superhorizon evolution of the curvature perturbation may be observable. While we hope to explore these possibilities in future work, we believe that without understanding the complete evolution of perturbations until they become adiabatic, it is somewhat premature to claim observable non-gaussianity as a prediction of multifield inflation.

The structure of the paper is as follows:  We begin in section \ref{themodel} with a discussion of the $\delta N$ formalism and the details of the model that we consider.  In section \ref{results} we give the results for the two-, three-, and four-point statistics.  In section \ref{fnlfate} we impose the condition that non-adiabatic fluctuations are damped away by a phase of effectively single field inflation, and we follow the evolution of $\tau_{\mathrm{NL}}$ and $g_{\mathrm{NL}}$ during this process.  We discuss the generalization of our work to $n$-point functions in section \ref{extensions}.  We conclude in section \ref{conclusion}.  Appendix \ref{details} gives the details for the calculation of three and four-point statistics, and appendix \ref{fiveandsix} gives explicit expressions for non-linearity parameters describing five and six-point functions.

\section{The Model}\label{themodel}
We use the same model as in \cite{Meyers:2010rg}, and, as in that work, we closely follow the treatment of Vernizzi and Wands \cite{Vernizzi:2006ve} (see also \cite{Seery:2005gb}) in our use of the $\delta N$ formalism. What follows is a somewhat condensed version of the derivations in \cite{Meyers:2010rg} -- the interested reader should see that work for details.

We work with canonical two field models of the form:
\begin{equation}
S = \int\textrm{d}^4x\sqrt{-g}\left[\frac{1}{2}m_p^2R+\frac{1}{2}g^{\mu\nu}\delta_{ab}\partial_{\mu}\phi^a\partial_{\nu}\phi^b-W(\vec{\phi})\right] \, .
\end{equation}
Using $\phi$ and $\chi$ for the two scalar fields, the slow roll equations of motion are (assuming $|\dot{H}|\ll H^2$ and $|\ddot{\phi_a}|\ll H|\dot{\phi_a}|$):
\begin{align}\label{eqm1}
3H\dot{\phi}&\simeq-\partial_{\phi}W\, ,\nonumber\\
3H\dot{\chi}&\simeq-\partial_{\chi}W\, ,\nonumber\\
H^2&\simeq\frac{1}{3m_p^2}W\, .
\end{align}
The validity of slow roll throughout superhorizon evolution is open to question. However, as argued in \cite{Meyers:2010rg}, brief violations of slow roll should not substantially change the results.

Because of their relative tractability, models of this type and their associated non-gaussianities have been well studied in the literature, for specific classes of potentials -- for example sum-separable potentials (e.g. \cite{Vernizzi:2006ve}), product-separable potentials (e.g. \cite{GarciaBellido:1995qq,Choi:2007su}) and many others (e.g. \cite{Wands:2010af,Byrnes:2009qy,Battefeld:2006sz,Chen:2009we,Sasaki:2008uc,Gao:2008dt,Battefeld:2009ym,Byrnes:2008zy}).

As alluded to above, studying such a system is most readily done using the $\delta N$ formalism \cite{Starobinsky:1986fxa,Sasaki:1995aw,Lyth:2004gb,Lyth:2005fi} to calculate the evolution of the curvature perturbation on uniform density hypersurfaces, $\zeta$ \cite{Bardeen:1980kt,Bardeen:1983qw}, which is conserved for adiabatic modes whose physical size is greater than that of the horizon.
 The $\delta N$ formalism relates $\zeta$ at some comoving time $t_c$ to the perturbation in the number of e-foldings from an initially flat hypersurface at $t=t_*$, generally taken to be horizon exit:
\begin{equation}\label{Nzeta}
\zeta\left(t_c,\vec{x}\right)\simeq\delta N\left(t_c,t_*,\vec x\right)\equiv\mathcal{N}\left(t_c,t_*,\vec x\right)-N\left(t_c,t_*\right)\, .
\end{equation}
$N$ is the unperturbed number of e-foldings, given by integrating $H$ from $t_*$ to $t_c$:
\begin{equation}\label{Nint}
N=\int_*^cH\textrm{d}t \, .
\end{equation}
Viewing the number of e-foldings as a function of the field configuration on the hypersurface defined by horizon exit, $\phi^I(t_*,\vec{x})$, and of $t_c$, the perturbation in $N$ can then be expressed in terms of the fluctuations of the scalar fields at horizon exit, shown here to third order:
\begin{equation}\label{deltaN}
\delta N\simeq\sum_I N_{,I}\delta\phi_*^I+\sum_{IJ}N_{,IJ}\delta\phi_*^I\delta\phi_*^J+\sum_{IJK}N_{,IJK}\delta\phi_*^I\delta\phi_*^J\delta\phi_*^K\, .
\end{equation}
The derivatives here are taken with respect to the fields at $t=t_*$ ($N_{,I} \equiv \frac{\partial N}{\partial \phi_*^I}$) and $\delta N$ is independent of the initial velocities since the potential is assumed to be slow roll at horizon exit.

\subsubsection*{Statistics}
From the above formalism, one can straightforwardly derive expressions for cosmological observables in terms of derivatives of the number of e-foldings. As in \cite{Meyers:2010rg} we follow the treatment of Vernizzi and Wands \cite{Vernizzi:2006ve} and in addition we also use the work of Byrnes, Sasaki and Wands \cite{Byrnes:2006vq} -- for a more detailed exposition of the derivation of $\delta N$ statistics, please see those works. We begin by defining the curvature and scalar power spectra, $\mathcal{P}_{\zeta}$ and $\mathcal{P}_*$:
\begin{align}\label{Pzetastar}
\left\langle\zeta_{\boldsymbol{k}_1}\zeta_{\boldsymbol{k}_2}\right\rangle&\equiv(2\pi)^3\delta^{(3)}(\boldsymbol{k}_1+\boldsymbol{k}_2)\frac{2\pi^2}{k_1^3}\mathcal{P}_{\zeta}(k_1)\, ,\nonumber\\
\left\langle\delta\phi^I_{\boldsymbol{k}_1}\delta\phi^J_{\boldsymbol{k}_2}\right\rangle&\equiv(2\pi)^3\delta^{IJ}\delta^{(3)}(\boldsymbol{k}_1+\boldsymbol{k}_2)\frac{2\pi^2}{k_1^3}\mathcal{P}_*(k_1)\, , \nonumber\\
P_*(k)&\equiv\frac{H_*^2}{4\pi^2}\, .
\end{align}
Then, from (\ref{Nzeta}), (\ref{deltaN}) and (\ref{Pzetastar}), we have:
\begin{equation}\label{PN}
\mathcal{P}_{\zeta}=\sum_IN_{,I}^2\mathcal{P}_*\, .
\end{equation}
The spectral index is given by (the approximate equality denotes lowest order in slow roll):
\begin{equation}\label{ns1}
n_{\zeta}-1\simeq\frac{\textrm{d}\ln\mathcal{P}_{\zeta}}{\textrm{d}N}=-2\epsilon+\frac{2}{H}\frac{\sum_{IJ}\dot{\phi}_JN_{,IJ}N_{,I}}{\sum_KN^2_{,K}}\, .
\end{equation}
$\epsilon\equiv-\dot{H}/H^2$ is the usual slow roll parameter. Using the slow roll equations of motion, this can be written as \cite{Lyth:1998xn}:
\begin{equation}\label{ns2}
n_{\zeta}-1=-2\epsilon-\frac{2}{m_p^2\sum_KN^2_{,K}}+\frac{2m_p^2\sum_{IJ}W_{,IJ}N_{,I}N_{,J}}{W\sum_KN^2_{,K}}\, .
\end{equation}

Three point statistics can be obtained in much the same fashion. The curvature bispectrum, $B_{\zeta}$, is defined through:
\begin{equation}\label{Bzeta}
\left\langle\zeta_{\boldsymbol{k}_1}\zeta_{\boldsymbol{k}_2}\zeta_{\boldsymbol{k}_3}\right\rangle\equiv(2\pi)^3\delta^{(3)}\left(\sum_i\boldsymbol{k}_i\right)\mathcal{B}_{\zeta}(k_1,k_2,k_3)\, .
\end{equation}
The bispectrum can be used to define the non-linearity parameter $f_{\mathrm{NL}}$ \cite{Komatsu:2001rj} which is the quantity most often referenced in observational constraints:
\begin{equation}\label{fnldef}
\frac{6}{5}f_{\mathrm{NL}}\equiv\frac{\prod_ik_i^3}{\sum_ik_i^3}\frac{B_{\zeta}}{4\pi^4\mathcal{P}_{\zeta}^2}\, .
\end{equation}
From (\ref{Nzeta}) and (\ref{deltaN}):
\begin{widetext}
\begin{multline}\label{3ptN}
\left\langle\zeta_{\boldsymbol{k}_1}\zeta_{\boldsymbol{k}_2}\zeta_{\boldsymbol{k}_3}\right\rangle=\sum_{IJK}N_{,I}N_{,J}N_{,K}\left\langle\delta\phi^I_{\boldsymbol{k}_1}\delta\phi^J_{\boldsymbol{k}_2}\delta\phi^K_{\boldsymbol{k}_3}\right\rangle+\frac{1}{2}\sum_{IJKL}N_{,I}N_{,J}N_{,KL}\left\langle\delta\phi^I_{\boldsymbol{k}_1}\delta\phi^J_{\boldsymbol{k}_2}(\delta\phi^K\star\delta\phi^L)_{\boldsymbol{k}_3}\right\rangle+\textrm{perms}\, .
\end{multline}
\end{widetext}
The star denotes a convolution and higher order terms have been neglected \cite{Zaballa:2006pv,Seery:2005gb}.

With some work (see \cite{Meyers:2010rg} for a more detailed exposition):
\begin{multline}
B(k_1,k_2,k_3)=4\pi^4\mathcal{P}_{\zeta}^2\frac{\sum_ik_i^3}{\prod_ik_i^3}\\
\times\left(\frac{-1}{4m_p^2\sum_KN_{,K}^2}\frac{\mathcal{F}}{\sum_ik_i^3}+\frac{\sum_{IJ}N_{,I}N_{,J}N_{,IJ}}{\left(\sum_KN_{,K}^2\right)^2}\right)\, ,
\end{multline}
with:
\begin{multline}
\mathcal{F}(k_1,k_2,k_3)=\\
-2\left(\frac{1}{2}\sum_{i\neq j}k_ik_j^2+4\frac{\sum_{i>j}k_i^2k_j^2}{k_t}-\frac{1}{2}\sum_ik_i^3\right)\, .
\end{multline}
From the above and (\ref{fnldef}), $f_{\mathrm{NL}}$ is:
\begin{equation}\label{fnltotal}
\frac{6}{5}f_{\mathrm{NL}}=\frac{\mathcal{P}_*}{2m_p^2\mathcal{P}_{\zeta}}(1+f)+\frac{\sum_{IJ}N_{,I}N_{,J}N_{,IJ}}{(\sum_KN_{,K}^2)^2}\, .
\end{equation}
The function $f\left(\equiv-1-\frac{\mathcal{F}}{2\sum_ik_i^3}\right)$ takes values between 0 and 5/6 and depends upon the shape of the bispectrum \cite{Maldacena:2002vr}. The first term of (\ref{fnltotal}) is proportional to the tensor-to-scalar ratio, $r$, and so is guaranteed to be small \cite{Komatsu:2010fb}, thus only the second term can give rise to a large $f_{\mathrm{NL}}$:
\begin{equation}
\frac{6}{5}f_{\mathrm{NL}}^{(4)}\equiv\frac{\sum_{IJ}N_{,I}N_{,J}N_{,IJ}}{(\sum_KN_{,K}^2)^2}\, .
\end{equation}
This term, $f_{\mathrm{NL}}^{(4)}$, is momentum-independent and local in real space, so it contributes to $f_{\mathrm{NL}}^{\mathrm{local}}$, which, as mentioned in the introduction, can be a sharp probe of the number of dynamical degrees of freedom during inflation. However, as we have shown in \cite{Meyers:2010rg}, generating and keeping a large $f_{\mathrm{NL}}^{\textrm{local}}$ along with an adiabatic power spectrum is far from easy.

The leading contribution to the 4-point function is given by \cite{Byrnes:2006vq}:
\begin{widetext}
\begin{align}\label{4pt}
\left\langle\zeta_{\boldsymbol{k}_1}\zeta_{\boldsymbol{k}_2}\zeta_{\boldsymbol{k}_3}\zeta_{\boldsymbol{k}_4}\right\rangle_c&=N_{,A}N_{,B}N_{,C}N_{,D}\left\langle\delta\phi^A_{\boldsymbol{k}_1}\delta\phi^B_{\boldsymbol{k}_2}\delta\phi^C_{\boldsymbol{k}_3}\delta\phi^D_{\boldsymbol{k}_4}\right\rangle_c\nonumber\\
&+\frac{1}{2}N_{,A_1A_2}N_{,B}N_{,C}N_{,D}\left[\left\langle\left(\delta\phi^{A_1}\star\delta\phi^{A_2}\right)_{\boldsymbol{k}_1}\delta\phi^B_{\boldsymbol{k}_2}\delta\phi^C_{\boldsymbol{k}_3}\delta\phi^D_{\boldsymbol{k}_4}\right\rangle+(\textrm{3 perms})\right]\nonumber\\
&+\frac{1}{2}N_{,A_1A_2}N_{,B_1B_2}N_{,C}N_{,D}\left[\left\langle\left(\delta\phi^{A_1}\star\delta\phi^{A_2}\right)_{\boldsymbol{k}_1}\left(\delta\phi^{B_1}\star\delta\phi^{B_2}\right)_{\boldsymbol{k}_2}\delta\phi^C_{\boldsymbol{k}_3}\delta\phi^D_{\boldsymbol{k}_4}\right\rangle+(\textrm{5 perms})\right]\nonumber\\
&+\frac{1}{2}N_{,A_1A_2A_3}N_{,B}N_{,C}N_{,D}\left[\left\langle\left(\delta\phi^{A_1}\star\delta\phi^{A_2}\star\delta\phi^{A_3}\right)_{\boldsymbol{k}_1}\delta\phi^B_{\boldsymbol{k}_2}\delta\phi^C_{\boldsymbol{k}_3}\delta\phi^D_{\boldsymbol{k}_4}\right\rangle+(\textrm{3 perms})\right]\, .
\end{align}
\end{widetext}
The subscript $c$ denotes the connected part of the trispectrum; there is also a disconnected contribution (consisting of parallelograms of wavevectors) present even for gaussian fields. The trispectrum, $T_{\zeta}(\boldsymbol{k}_1,\boldsymbol{k}_2,\boldsymbol{k}_3,\boldsymbol{k}_4)$ can then be defined \cite{Byrnes:2006vq} as:
\begin{equation}
\left\langle\zeta_{\boldsymbol{k}_1}\zeta_{\boldsymbol{k}_2}\zeta_{\boldsymbol{k}_3}\zeta_{\boldsymbol{k}_4}\right\rangle_c\equiv T_{\zeta}\left(\boldsymbol{k}_1,\boldsymbol{k}_2,\boldsymbol{k}_3,\boldsymbol{k}_4\right)\left(2\pi\right)^3\delta^3\left(\sum_i\boldsymbol{k}_i\right)\, .
\end{equation}

Although the general expression for $T_{\zeta}$ is quite involved, it simplifies considerably in the case where we limit ourselves to models with independent gaussian fluctuations at horizon exit \cite{Byrnes:2006vq}:
\begin{align}
T_{\zeta}(\boldsymbol{k}_1,\boldsymbol{k}_2,\boldsymbol{k}_3,\boldsymbol{k}_4)&=\tau_{\mathrm{NL}}\left[P_{\zeta}(k_{13})P_{\zeta}(k_{3})P_{\zeta}(k_{4})+\textrm{(11 perms)}\right]\nonumber\\
&+\frac{54}{25}g_{\mathrm{NL}}\left[P_{\zeta}(k_{2})P_{\zeta}(k_{3})P_{\zeta}(k_{4})+\textrm{(3 perms)}\right]\, .
\end{align}
$k_{ij}=|\boldsymbol{k}_i+\boldsymbol{k}_j|$, and $\tau_{\mathrm{NL}}$ and $g_{\mathrm{NL}}$ are given by:
\begin{align}
\tau_{\mathrm{NL}}&=\frac{\sum_{IJK}N_{,IJ}N_{,IK}N_{,J}N_{,K}}{\left(\sum_LN^2_{,L}\right)^3}\, ,\nonumber\\
g_{\mathrm{NL}}&=\frac{25}{54}\frac{\sum_{IJK}N_{,IJK}N_{,I}N_{,J}N_{,K}}{\left(\sum_LN^2_{,L}\right)^3}\, .
\end{align}
Note that the parametrization is such that the coefficients multiply functions with different $k$-dependence, making them observationally distinguishable. Restricting oneself to the above form gives the local-form non-gaussianity of the trispectrum.

Before moving on we should point out that much of the recent interest in the trispectrum is due to the discovery of a useful inequality by Suyama and Yamaguchi, relating $f_{\mathrm{NL}}^{(4)}$ to $\tau_{\mathrm{NL}}$ \cite{Suyama:2007bg,Sugiyama:2011jt}:
\begin{equation}
\tau_{\mathrm{NL}}\geq\left(\frac{6f_{\mathrm{NL}}^{(4)}}{5}\right)^2\, .
\end{equation}
This is derived from the Cauchy-Schwarz inequality
\begin{equation}
\left(\sum_Ia_I^2\right)\left(\sum_Jb_J^2\right)\geq\left(\sum_Ia_Ib_I\right)^2
\end{equation}
with
\begin{align}
a_I&=\frac{\sum_JN_{,IJ}N_{,J}}{\left[\sum_KN^2_{,K}\right]^{3/2}}\, ,\nonumber\\
b_I&=\frac{N_{,I}}{\left[\sum_KN^2_{,K}\right]^{1/2}}\, ,
\end{align}
Current experimental bounds are consistent with this inequality, but should it be violated in future measurements, a large class of multifield models would be ruled out. Of course, such an inequality would be of less utility if the approach to adiabaticity wipes out any possibility of an observable (local) non-gaussian signal.

\subsubsection*{The Potential}\label{potentialdetails}
Requiring that the $\delta N$ formalism be of practical use places restrictions on the form of the potential, discussed in detail in \cite{Meyers:2010rg}. The principal point is that in order to take derivatives of $N$ with respect to the initial field values we need a way to easily relate the final and initial field values. This can be done if a suitable constant of motion can be constructed, which restricts the form of the potential to:
\begin{equation}
W(\phi,\chi)\equiv F(U(\phi)+V(\chi))\, ,
\end{equation}
and also requires that there is a one-to-one mapping between $\phi$ and $\chi$. With $W$ of this form, the appropriate constant of motion is:
\begin{equation}\label{Cpotential}
C=-m_p^2\int_{\phi_0}^{\phi}\frac{1}{U'(\phi')}\textrm{d}\phi'+m_p^2\int_{\chi_0}^{\chi}\frac{1}{V'(\chi')}\textrm{d}\chi'\, .
\end{equation}
The slow roll equations of motion are:
\begin{align}
3H\dot{\phi}&\simeq F'U'\, ,\nonumber\\
3H\dot{\chi}&\simeq F'V'\, ,\nonumber\\
H^2&\simeq\frac{1}{3m_p^2}W\left(\phi,\chi\right)\, .
\end{align}
Here, as below, a prime denotes the derivative with respect the argument of the function. The slow roll parameters $\epsilon^I=(m_p^2/2)(\partial W/\partial\phi_I)^2$ and $\eta^{IJ}=m_p^2(\partial^2W/\partial\phi_I\partial\phi_J)$ are given by (we drop the repeated index on the diagonal $\eta$ for conciseness):
\begin{align}\label{slowroll}
\epsilon^{\phi}&\equiv\frac{m_p^2}{2}\left(\frac{F'U'}{F}\right)^2\, ,\nonumber\\
\epsilon^{\chi}&\equiv\frac{m_p^2}{2}\left(\frac{F'V'}{F}\right)^2\, ,\nonumber\\
\epsilon&=\epsilon^{\phi}+\epsilon^{\chi}=-\frac{\dot{H}}{H^2}\, ,\nonumber\\
\eta^{\phi}&\equiv m_p^2\left(\frac{F''U'^2+F'U''}{F}\right)\, ,\nonumber\\
\eta^{\chi}&\equiv m_p^2\left(\frac{F''V'^2+F'V''}{F}\right)\, ,\nonumber\\
\eta^{\phi\chi}&\equiv m_p^2\left(\frac{F''U'V'}{F}\right)=2\frac{FF''}{F'^2}\sqrt{\epsilon^{\phi}\epsilon^{\chi}}\, .
\end{align}

With the potential given above, $N$ is now of the form:
\begin{multline}\label{Npotential}
N=-\frac{1}{2m_p^2}\int_{*}^{c}\frac{W\left(\phi,\chi(\phi)\right)}{W_{\phi}\left(\phi,\chi(\phi)\right)}\textrm{d}\phi\\
-\frac{1}{2m_p^2}\int_{*}^{c}\frac{W\left(\phi(\chi),\chi\right)}{W_{\chi}\left(\phi(\chi),\chi\right)}\textrm{d}\chi\, .
\end{multline}
We impose a further restriction by requiring that the integral defining $N$ can be divided into separate integrals over $\phi$ and $\chi$, without using the functions $\phi(\chi)$ or $\chi(\phi)$. This gives two possibilities for the potential homogeneous
\begin{equation}
W=\left[U(\phi)+V(\chi)\right]^{\gamma}\, ,
\end{equation}
and exponential
\begin{equation}\label{exppot}
W=W_0\,\textrm{Exp}\left[U(\phi)+V(\chi)\right]\, .
\end{equation}
For the homogeneous case, $N$ is then given by:
\begin{equation}\label{Nhomogeneous}
N=-\frac{1}{\gamma m_p^2}\int_*^c\frac{U(\phi)}{U'\left(\phi\right)}\textrm{d}\phi
-\frac{1}{\gamma m_p^2}\int_*^c\frac{V(\chi)}{V'\left(\chi\right)}\textrm{d}\chi\, .
\end{equation}
And for the exponential case:
\begin{equation}\label{Nexpo}
N=-\frac{1}{2m_p^2}\int_*^c\frac{1}{U'\left(\phi\right)}\textrm{d}\phi
-\frac{1}{2m_p^2}\int_*^c\frac{1}{V'\left(\chi\right)}\textrm{d}\chi\, .
\end{equation}

When we end inflation we will also be interested in the parameters $\eta^{\sigma\sigma}$ and $\eta^{ss}$, where $\sigma$ and $s$ refer the directions parallel and perpendicular to the inflaton motion in field space.  These parameters control the masses of the adiabatic and isocurvature perturbation, respectively. In section \ref{fnlfate}, we will be interested in the case where the universe passes through a phase of single field inflation where the isocurvature perturbations become heavy while the adiabatic ones stay light. These parameters are given by \cite{Gordon:2000hv}:
\begin{align}\label{etass1}
\eta^{\sigma\sigma}&\equiv\frac{\epsilon^{\phi}\eta^{\phi}+2\sqrt{\epsilon^{\phi}\epsilon^{\chi}}\eta^{\phi\chi}+\epsilon^{\chi}\eta^{\chi}}{\epsilon}\nonumber\\
\eta^{ss}&\equiv\frac{\epsilon^{\chi}\eta^{\phi}-2\sqrt{\epsilon^{\phi}\epsilon^{\chi}}\eta^{\phi\chi}+\epsilon^{\phi}\eta^{\chi}}{\epsilon} \, .
\end{align}

Before moving on to our results and discussion thereof, we note that although our calculations assume slow roll, small deviations are possible. In particular, violations of slow roll are consistent with our $\delta N$ analysis so long as either the violation is sufficiently short or if $H/\dot{\phi}$ is only weakly-dependent on the initial field value (attractive trajectory) and friction is subdominant during any non-slow roll regime. The details of this argument can be found in \cite{Meyers:2010rg}. There is also some numerical evidence to suggest the applicability of $\delta N$ beyond the slow roll regime -- see \cite{Vernizzi:2006ve} and \cite{Wanatanabe:2009}.

\section{Results}\label{results}
We include a summary of the results for both the homogeneous and exponential potentials here. The details of the calculation are presented in appendix A of \cite{Meyers:2010rg}. We reproduce that exposition and expand it to include four-point statistics in appendix \ref{details} below.

\subsubsection*{Homogeneous Potential: $W(\phi,\chi)=\left[U(\phi)+V(\chi)\right]^{\gamma}$}
Results here are expressed in terms of derivatives of the potential and the slow roll parameters defined in (\ref{slowroll}). As before a prime denotes a derivative with respect to the argument of the function -- $\phi$ for $U$ and $\chi$ for $V$.  First, the mass parameters for the adiabatic and isocurvature fluctuations are given by:
\begin{align}\label{etass2}
\eta^{\sigma\sigma}&=\frac{\epsilon^{\phi}\eta^{\phi}+4\frac{(\gamma-1)}{\gamma}\epsilon^{\phi}\epsilon^{\chi}+\epsilon^{\chi}\eta^{\chi}}{\epsilon}\, ,\nonumber\\
\eta^{ss}&=\frac{\epsilon^{\chi}\eta^{\phi}-4\frac{(\gamma-1)}{\gamma}\epsilon^{\phi}\epsilon^{\chi}+\epsilon^{\phi}\eta^{\chi}}{\epsilon}\, .
\end{align}

For the other quantities of interest, we first define:
\begin{align}\label{xy}
x_h&\equiv\frac{1}{U_*+V_*}\left(U_*+\frac{V_c\epsilon_c^{\phi}-U_c\epsilon_c^{\chi}}{\epsilon_c}\right)\, ,\nonumber\\
y_h&\equiv\frac{1}{U_*+V_*}\left(V_*-\frac{V_c\epsilon_c^{\phi}-U_c\epsilon_c^{\chi}}{\epsilon_c}\right)\, .
\end{align}
Then the observable $\mathcal{P}_{\zeta}$ is given by
\begin{equation}\label{Pandetahomogeneous}
\mathcal{P}_{\zeta}=\frac{W_*}{24\pi^2m_p^4}\left(\frac{x_h^2}{\epsilon_*^{\phi}}+\frac{y_h^2}{\epsilon_*^{\chi}}\right)\, ,
\end{equation}
and $n_{\zeta}-1$ by
\begin{multline}
\label{nshomo}
n_{\zeta}-1=-2\epsilon_*-\frac{4}{\gamma}\left(\frac{x_h^2}{\epsilon_*^{\phi}}+\frac{y_h^2}{\epsilon_*^{\chi}}\right)^{-1}\\
\times\left(x_h\left[1-\left(\frac{\gamma\eta^{\phi}_*}{2\epsilon_*^{\phi}}-\gamma+1\right)x_h\right]\right.\\
+y_h\left[1-\left(\frac{\gamma\eta^{\chi}_*}{2\epsilon_*^{\chi}}-\gamma+1\right)y_h\right]\Bigg)\, .
\end{multline}

The full expressions for $f_{\mathrm{NL}}$, $\tau_{\mathrm{NL}}$ and $g_{\mathrm{NL}}$ are given in appendix \ref{details}. Many of the terms in these expressions, however, are multiplied by slow roll parameters. If we keep only those parts that are leading order in slow roll, we have (here and below we use $\varepsilon$ to refer to a generic first order slow roll parameter):
\begin{align}
\frac{6}{5}f_{\mathrm{NL}}^{(4)}&\sim\mathcal{O}\left(\varepsilon_*\right)+\frac{2}{\gamma}\left(\frac{\frac{\left(U_c+V_c\right)^2}{\left(U_*+V_*\right)^2}\left(\frac{x_h}{\epsilon_*^{\phi}}-\frac{y_h}{\epsilon_*^{\chi}}\right)^2\frac{\epsilon_c^{\phi}\epsilon_c^{\chi}}{\epsilon_c}\left(\frac{\gamma\eta_c^{ss}}{\epsilon_c}-1\right)}{\left(\frac{x_h^2}{\epsilon_*^{\phi}}+\frac{y_h^2}{\epsilon_*^{\chi}}\right)^2}\right)\, ,\label{fnlhomogeneous}
\end{align}
\begin{align}
\tau_{\mathrm{NL}}&\sim\mathcal{O}\left(\varepsilon_*\right)+\frac{4}{\gamma^2}\frac{\left(\frac{x_h}{\epsilon_*^{\phi}}-\frac{y_h}{\epsilon_*^{\chi}}\right)^2\left(\frac{1}{\epsilon_*^{\phi}}+\frac{1}{\epsilon_*^{\chi}}\right)}{\left(\frac{x_h^2}{\epsilon_*^{\phi}}+\frac{y_h^2}{\epsilon_*^{\chi}}\right)^3}\nonumber\\
&\qquad\qquad\qquad\times\left[\frac{\left(U_c+V_c\right)^2}{\left(U_*+V_*\right)^2}\frac{\epsilon_c^{\phi}\epsilon_c^{\chi}}{\epsilon_c}\left(\frac{\gamma\eta^{ss}_c}{\epsilon_c}-1\right)\right]^2\, , \label{taunlhomogeneous}
\end{align}
\begin{widetext}
\begin{multline}\label{gnlhomogeneous}
g_{\mathrm{NL}}\sim\mathcal{O}\left(\varepsilon_*\right)+\frac{50}{27}\frac{1}{\gamma}\frac{\left(\frac{x_h}{\epsilon_*^{\phi}}-\frac{y_h}{\epsilon_*^{\chi}}\right)^3}{\left(\frac{x_h^2}{\epsilon_*^{\phi}}+\frac{y_h^2}{\epsilon_*^{\chi}}\right)^3}\frac{\left(U_c+V_c\right)^3}{\left(U_*+V_*\right)^3}\frac{\epsilon_c^{\phi}\epsilon_c^{\chi}}{\epsilon_c^4}\\
\times\left[-\epsilon\left.\epsilon^{\chi}\right.^2\eta^{\phi}+\epsilon\left.\epsilon^{\phi}\right.^2\eta^{\chi}+\frac{2\left(\gamma-1\right)}{\gamma}\epsilon^{\phi}\epsilon^{\chi}\left(\left.\epsilon^{\chi}\right.^2-\left.\epsilon^{\phi}\right.^2\right)+\left.\epsilon^{\chi}\right.^2\eta^{\phi}\left(\gamma\eta^{\phi}-\frac{7\left(\gamma-1\right)}{\gamma}\epsilon^{\phi}\right)-\left.\epsilon^{\phi}\right.^2\eta^{\chi}\left(\gamma\eta^{\chi}-\frac{7\left(\gamma-1\right)}{\gamma}\epsilon^{\chi}\right)\right.\\
\left.+\frac{\gamma}{2}\left(\left.\epsilon^{\chi}\right.^2\left.\xi^{\phi}\right.^2-\left.\epsilon^{\phi}\right.^2\left.\xi^{\chi}\right.^2\right)+3\gamma\epsilon^{\phi}\epsilon^{\chi}\eta^{ss}\left(\eta^{\chi}-\frac{2\left(\gamma-1\right)}{\gamma}\epsilon^{\chi}-\eta^{\phi}+\frac{2\left(\gamma-1\right)}{\gamma}\epsilon^{\phi}\right)\right]_c\, .
\end{multline}
The $\xi$ are defined by:
\begin{align}\label{xi}
\left.\xi^{\phi}\right.^2&\equiv m_p^4\frac{W_{,\phi}W_{,\phi\phi\phi}}{W^2}=m_p^4\frac{\gamma^2\left(U+V\right)^2U'U'''+3\gamma^2\left(\gamma-1\right)\left(U+V\right)U'^2U''+\gamma^2\left(\gamma-1\right)\left(\gamma-2\right)U'^4}{\left(U+V\right)^4}\, ,\nonumber\\
\left.\xi^{\chi}\right.^2&\equiv m_p^4\frac{W_{,\chi}W_{,\chi\chi\chi}}{W^2}=m_p^4\frac{\gamma^2\left(U+V\right)^2V'V'''+3\gamma^2\left(\gamma-1\right)\left(U+V\right)V'^2V''+\gamma^2\left(\gamma-1\right)\left(\gamma-2\right)V'^4}{\left(U+V\right)^4}\, .
\end{align}
\end{widetext}

\subsubsection*{Exponential Potential: $W(\phi,\chi)=W_0\textrm{Exp}\left[U(\phi)+V(\chi)\right]$}
The slow roll parameters (\ref{slowroll}) take a particularly simple form for an exponential potential of the form (\ref{exppot}), and readily lead to the following expressions for the mass parameters $\eta^{ss}$ and $\eta^{\sigma\sigma}$:
\begin{align}\label{etassexp}
\eta^{\sigma\sigma}&\equiv\frac{\epsilon^{\phi}\eta^{\phi}+4\epsilon^{\phi}\epsilon^{\chi}+\epsilon^{\chi}\eta^{\chi}}{\epsilon}\, ,\nonumber\\
\eta^{ss}&\equiv\frac{\epsilon^{\chi}\eta^{\phi}-4\epsilon^{\phi}\epsilon^{\chi}+\epsilon^{\phi}\eta^{\chi}}{\epsilon} \, .
\end{align}

Next, in a similar fashion to the previous section, we define:
\begin{align}\label{xye}
x_e&\equiv \frac{2\epsilon_c^{\phi}}{\epsilon_c}\, , & y_e&\equiv \frac{2\epsilon_c^{\chi}}{\epsilon_c}\, .
\end{align}
Then the observables $\mathcal{P}_{\zeta}$ and $n_{\zeta}-1$ are given by:
\begin{align}\label{Pandnsexp}
\mathcal{P}_{\zeta}&=\frac{W_*}{96\pi^2m_p^4}\left(\frac{x_e^2}{\epsilon_*^{\phi}}+\frac{y_e^2}{\epsilon_*^{\chi}}\right)\, ,
\end{align}
\begin{align}
n_{\zeta}-1&=-2\epsilon_*-4\left(\frac{2-\left(\frac{\eta_*^{\phi}x_e^2}{2\epsilon_*^{\phi}}+2x_ey_e+\frac{\eta_*^{\chi}y_e^2}{2\epsilon_*^{\chi}}\right)}{\frac{x_e^2}{\epsilon_*^{\phi}}+\frac{y_e^2}{\epsilon_*^{\chi}}}\right)\, .
\end{align}
As before full expressions for $f_{\mathrm{NL}}$, $\tau_{\mathrm{NL}}$ and $g_{\mathrm{NL}}$ can be found in appendix \ref{details}, but if we keep only the leading terms in slow roll, we have:
\begin{widetext}
\begin{equation}
\label{fnlexp}
\frac{6}{5}f_{\mathrm{NL}}^{(4)}\sim\mathcal{O}\left(\varepsilon_*\right)+2\frac{\left(\frac{x_e}{\epsilon_*^{\phi}}-\frac{y_e}{\epsilon_*^{\chi}}\right)^2}{\left(\frac{x_e^2}{2\epsilon_*^{\phi}}+\frac{y_e^2}{2\epsilon_*^{\chi}}\right)^2}\frac{\epsilon_c^{\phi}\epsilon_c^{\chi}}{\epsilon_c^2}\eta_c^{ss}\, ,
\end{equation}
\begin{equation}
\label{taunlexp}
\tau_{\mathrm{NL}}\sim\mathcal{O}\left(\varepsilon_*\right)+\frac{64\left(y_e\epsilon_*^{\phi}-x_e\epsilon_*^{\chi}\right)^2\left(\epsilon_*^{\phi}+\epsilon_*^{\chi}\right)}{\left(y_e^2\epsilon_*^{\phi}+x_e^2\epsilon_*^{\chi}\right)^3}\left(\frac{\epsilon_c^{\chi}\epsilon_c^{\phi}}{\epsilon_c^2}\right)^2\left.\eta_c^{ss}\right.^2\, ,
\end{equation}
\begin{multline}
\label{gnlexp}
g_{\mathrm{NL}}\sim\mathcal{O}\left(\varepsilon_*\right)+\frac{200}{27}\frac{\left(y_e\epsilon_*^{\phi}-x_e\epsilon_*^{\chi}\right)^3}{\left(y_e^2\epsilon_*^{\phi}+x_e^2\epsilon_*^{\chi}\right)^3}\frac{\epsilon_c^{\chi}\epsilon_c^{\phi}}{\epsilon_c^5}\Bigg(\epsilon_c^{\phi}\left.\epsilon_c^{\chi}\right.^2\left(26\epsilon_c^{\chi}\eta_c^{\phi}+4\left.\eta_c^{\phi}\right.^2-6\eta_c^{\phi}\eta_c^{\chi}-\left.\xi_c^{\phi}\right.^2\right)-\left.\epsilon_c^{\chi}\right.^3\left(2\left.\eta_c^{\phi}\right.^2+\left.\xi_c^{\phi}\right.^2\right)\\
+\left.\epsilon_c^{\phi}\right.^3\left(48\left.\epsilon_c^{\chi}\right.^2-26\epsilon_c^{\chi}\eta_c^{\chi}+2\left.\eta_c^{\chi}\right.^2+\left.\xi_c^{\chi}\right.^2\right)+\left.\epsilon_c^{\phi}\right.^2\epsilon_c^{\chi}\left(-48\left.\epsilon_c^{\chi}\right.^2+6\eta_c^{\phi}\eta_c^{\chi}-4\left.\eta_c^{\chi}\right.^2+22\epsilon_c^{\chi}\left(-\eta_c^{\phi}+\eta_c^{\chi}\right)+\left.\xi_c^{\chi}\right.^2\right)\Bigg)\, .
\end{multline}
\end{widetext}

\section{Damping Away Isocurvature and the Fate of Non-Gaussianities}\label{fnlfate}
\subsubsection*{Homogeneous Potential: $W(\phi,\chi)=\left[U(\phi)+V(\chi)\right]^{\gamma}$}
We are now in a position to assess the magnitude of $g_{\mathrm{NL}}$ and $\tau_{\mathrm{NL}}$ as isocurvature is damped away, beginning with the homogeneous potential. We are only interested in the evolution of non-slow roll suppressed parts of the local trispectrum, so we will focus on the leading terms given in (\ref{taunlhomogeneous}) and (\ref{gnlhomogeneous}).

In order to translate the correlation functions of primordial fluctuations into observations, we must follow their evolution until they become conserved quantities, or until they are observed. The latter would require detailed knowledge of the cosmological history from the present all the way back to the inflationary epoch. Given our ignorance of the details of much of the early universe, we focus on the former approach. Accordingly we wish to study $\tau_{\mathrm{NL}}$ and $g_{\mathrm{NL}}$ as they are forced into constant values and, since correlations of the curvature perturbation will evolve outside the horizon as long as non-adiabatic fluctuations are present, this means we need to evaluate (\ref{taunlhomogeneous}) and (\ref{gnlhomogeneous}) after any non-adiabaticity has been damped away and only the adiabatic mode is left. Here we achieve this by requiring that there is a short phase of effectively single field inflation sometime before reheating while the observationally relevant modes are outside the horizon. We repeat that this is not the only possibility for achieving adiabaticity, but this choice ensures that we are able to calculate observables without making any assumptions about post-inflationary dynamics. In \cite{Meyers:2010rg} we perform the same analysis and find that $f_{\mathrm{NL}}^{(4)}$ becomes slow roll suppressed as we approach adiabaticity.

We now consider the equation of motion for the isocurvature fluctuations which at leading order in slow roll takes the form \cite{Gordon:2000hv}:
\begin{equation}\label{nonadiabatic}
\ddot{\delta s} + 3H\dot{\delta s} + \frac{W}{m_P^2}\eta^{ss}\delta s + \frac{k^2}{a^2}\delta s = 0\, .
\end{equation}
The solution of this equation is \cite{Gordon:2000hv,Riotto:2002yw}
\begin{equation}\label{deltasgeneral}
\delta s \propto a(t)^{-3/2}\left(\frac{k}{aH}\right)^{-\nu}\, ,
\end{equation}
where $\nu$ is given by
\begin{equation}\label{nu}
\nu^2=\frac{9}{4}-\frac{W\eta^{ss}}{m_P^2H^2}\, .
\end{equation}	
So for $\eta^{ss} \geq \frac{3}{4}$, we find
\begin{align}\label{deltas}
|\delta s|&\propto a(t)^{-3/2}\nonumber\\
\Rightarrow|\delta s|&\sim \textrm{Exp}\left[-\frac{3}{2}\int H\textrm{d}t\right]\, ,
\end{align}
and the isocurvature fluctuations are rapidly damped away.

We will now examine the conditions for large $\eta^{ss}$, defined in (\ref{etass1}). Recall that (\ref{slowroll}) gives
\begin{equation}\label{etaphichi}
\eta^{\phi\chi}=2\frac{(\gamma-1)}{\gamma}\sqrt{\epsilon^{\phi}\epsilon^{\chi}}\, ,
\end{equation}
and (from (\ref{nshomo})) $|\gamma|$ must be $\mathcal{O}(1)$ or larger to guarantee scale invariance.  Thus when
\begin{equation}
\eta^{ss}=\frac{\epsilon^{\chi}\eta^{\phi}-4\frac{(\gamma-1)}{\gamma}\epsilon^{\phi}\epsilon^{\chi}+\epsilon^{\phi}\eta^{\chi}}{\epsilon}\, .\tag{\ref{etass2}}
\end{equation}
is large, we must have either $\eta^{\phi}\gtrsim\frac{\epsilon}{\epsilon^{\chi}}$ or $\eta^{\chi}\gtrsim\frac{\epsilon}{\epsilon^{\phi}}$.  If both $\eta^{\phi}$ and $\eta^{\chi}$  are large, then $\eta^{\sigma\sigma}$ will also be large, and inflation will quickly end before the non-adiabatic modes have been damped away, so we will not be interested in this case.

Now let us consider $\tau_{\mathrm{NL}}$, which we write in the following simplified form:
\begin{equation}
\tau_{\mathrm{NL}}\sim\mathcal{O}\left(\varepsilon_*\right)+\mathcal{O}\left(1\right)\times\left[\frac{\epsilon_c^{\phi}\epsilon_c^{\chi}}{\epsilon_c}\left(\frac{\gamma\eta_c^{ss}}{\epsilon_c}-1\right)\right]^2\, .
\end{equation}
Given the appearance of $\eta^{ss}$, one might expect that for sufficiently large $\eta^{ss}$ (and large $\eta^{ss}$ is required to damp away isocurvature) observably large $\tau_{\mathrm{NL}}$ could be produced even as isocurvature is damped away. However, the coefficient of $\eta^{ss}$, specifically $\frac{\epsilon^{\phi}\epsilon^{\chi}}{\epsilon^2}$, is a dynamical quantity whose time-dependence is affected by the behavior of $\eta^{ss}$.  To find this time-dependence, we first note that:
\begin{align}
\frac{\dot{\epsilon^\phi}}{H} &=-2\epsilon^\phi\eta^\phi-2\sqrt{\epsilon^\phi\epsilon^\chi}\eta^{\phi\chi}+4(\epsilon^\phi)^2+4\epsilon^\phi\epsilon^\chi \label{epsilonphidot}\\
\frac{\dot{\epsilon^\chi}}{H} &=-2\epsilon^\chi\eta^\chi-2\sqrt{\epsilon^\phi\epsilon^\chi}\eta^{\phi\chi}+4(\epsilon^\chi)^2+4\epsilon^\phi\epsilon^\chi \label{epsilonchidot}.
\end{align}
Now, for the damping of isocurvature either $\eta^{\phi}>1$ or $\eta^{\chi}>1$. In the former case we can neglect all but the first term on the right hand side of (\ref{epsilonphidot}), and so we find that
\begin{equation}\label{epsilonphidamping}
\epsilon^{\phi}(t) \propto \textrm{Exp}\left[-2\int H\eta^{\phi}\rm{d}t\right].
\end{equation}
Similar remarks apply to $\epsilon^{\chi}$, in the case that $\eta^{\chi}>1$.

It it is then straightforward to write down the evolution of $\tau_{\mathrm{NL}}$ as isocurvature is damped away:
\begin{align}\label{taudamping}
\tau_{\mathrm{NL}}&\sim\mathcal{O}(\varepsilon_*)+\mathcal{O}(1)\times\left(\eta^{ss}\right)^2\textrm{Exp}\left[-4\int C_{\eta}H\eta^{ss}\textrm{d}t\right]\, .
\end{align}
$C_{\eta}$ is a number which is always greater than 1 whose value depends on the particular direction of the effective single field during this phase. We thus conclude that $\tau_{\mathrm{NL}}$ will always be slow roll suppressed upon entering the purely adiabatic solution after a phase of effectively single field inflation.

The argument is similar for $g_{\mathrm{NL}}$.  By examining (\ref{gnlhomogeneous}), we can see that the part of $g_{\mathrm{NL}}$ which is not slow roll suppressed is proportional to $\frac{\epsilon^{\phi}\epsilon^{\chi}}{\epsilon^2}$.  As we argued above, either $\epsilon^{\phi}$ or $\epsilon^{\chi}$ will be exponentially damped when $\eta^{ss}$ becomes large.  In order for $g_{\mathrm{NL}}$ to remain large as non-adiabatic fluctuations are damped away, one of the terms in square brackets in (\ref{gnlhomogeneous}) must grow exponentially.  Since we must have $\epsilon<1$ throughout inflation, we cannot have exponential growth from terms which contain only $\epsilon^{\phi}$ and $\epsilon^{\chi}$.  An exponentially growing $\eta^{\phi}$ would result in a double exponential damping of $\epsilon^{\phi}$, as can be seen from (\ref{epsilonphidamping}), and thus a rapid damping of $g_{\mathrm{NL}}$ (and likewise for $\eta^{\chi}$).

This leaves only the terms proportional to $\xi^2$.  However, the evolution of $\xi^2$ is not independent of that of the other slow roll parameters.  Specifically, we note that
\begin{align}
\frac{\dot{\eta^{\phi}}}{H}&=-\left.\xi^{\phi}\right.^2+2\epsilon^{\phi}\eta^{\chi}+\frac{2}{\gamma}\epsilon^{\chi}\eta^{\phi}+\frac{2}{\gamma}\sqrt{\epsilon^{\phi}\epsilon^{\chi}}\eta^{\phi\chi} \label{etaphidot}\\
\frac{\dot{\eta^{\chi}}}{H}&=-\left.\xi^{\chi}\right.^2+2\epsilon^{\chi}\eta^{\phi}+\frac{2}{\gamma}\epsilon^{\phi}\eta^{\chi}+\frac{2}{\gamma}\sqrt{\epsilon^{\phi}\epsilon^{\chi}}\eta^{\phi\chi} \label{etachidot} \, .
\end{align}
If $\left.\xi^{\phi}\right.^2$ were positive and exponentially growing, this would mean that $\eta^{\phi}$ would quickly become negative and exponentially growing.  This would lead to a double exponential growth of $\epsilon^{\phi}$ which would result in a quick end to inflation before the non-adiabatic fluctuations are damped away.  If $\left.\xi^{\phi}\right.^2$ were negative and exponentially growing, then $\eta^{\phi}$ would exponentially increase as well, which in turn would result in a double exponential suppression of $\epsilon^{\phi}$, and thus a rapid damping of $g_{\mathrm{NL}}$. The same analysis holds for $\left.\xi^{\chi}\right.^2$.  So we find that like $f_{\mathrm{NL}}^{(4)}$ and $\tau_{\mathrm{NL}}$, $g_{\mathrm{NL}}$ will always be slow roll suppressed upon entering a purely adiabatic solution after a phase of effectively single field inflation.

We pause to note that while the terms labeled $\mathcal{O}(\varepsilon_*)$ are proportional to slow roll parameters, there are cases when they are not negligible.  As shown in Appendix \ref{details}, these terms contain several factors which depend on the details of the potential. In fact, there are examples showing that these terms can be significant for $f_{\mathrm{NL}}^{(4)}$ \cite{Byrnes:2008wi,Kim:2010ud}, and a similar mechanism should apply for higher point statistics as well.

\subsubsection*{Exponential Potential: $W(\phi,\chi)=W_0\textrm{Exp}\left[U(\phi)+V(\chi)\right]$}
An entirely analogous damping takes place in the case of the exponential potential. Once again the dominant contribution to $\tau_{\mathrm{NL}}$ will be proportional to $\left.\eta^{ss}_c\right.^2$
\begin{equation}
\tau_{\mathrm{NL}}\sim\mathcal{O}\left(\varepsilon_*\right)+\mathcal{O}(1)\times\left(\frac{\epsilon_c^{\chi}\epsilon_c^{\phi}}{\epsilon_c^2}\right)^2\left.\eta_c^{ss}\right.^2\, ,
\end{equation}
and $\eta^{ss}$ is
\begin{equation}
\eta^{ss}\equiv\frac{\epsilon^{\chi}\eta^{\phi}-4\epsilon^{\phi}\epsilon^{\chi}+\epsilon^{\phi}\eta^{\chi}}{\epsilon} \, .\tag{\ref{etassexp}}
\end{equation}
A large $\eta^{ss}$ will then imply (again) either a large $\eta^{\phi}$ or a large $\eta^{\chi}$ and lead (in the same fashion as for the homogeneous potential) to the exponential damping of the leading term in $\tau_{\mathrm{NL}}$ as in (\ref{taudamping}).

The argument for $g_{\mathrm{NL}}$ is similar. The exponentially suppressed factor $\frac{\epsilon^{\phi}\epsilon^{\chi}}{\epsilon}$ also multiplies the leading order term in $g_{\mathrm{NL}}$ and the factors in the expression cannot be exponentially large (by the same arguments as above). Thus the leading term in $g_{\mathrm{NL}}$ is driven to zero as the fluctuations are forced into an adiabatic mode.

\section{Higher Point Functions}\label{extensions}
Although we have examined only the three- and four-point functions in detail, we can extend our analysis to higher point functions as well.  The $n$-point function is always proportional to $n$ factors of $N$ with at least one derivative acting on each $N$.  If we restrict our attention to models where the scalar field fluctuations are gaussian and independent at horizon exit, then the contributions to the non-gaussian correlation functions are proportional to the various ways of contracting the derivatives of $N$ with Kronecker deltas.  If we focus on the connected part of the non-gaussian correlations, then there is no subset of the factors of $N$ whose derivatives are contracted independently of the other factors.  In this case the leading contributions to the $n$-point function will contain $2(n-1)$ derivatives.  We will briefly discuss the sub-leading contributions below.

For a general $n$-point function we can characterize the various local forms by defining a set of non-linearity parameters.  The number of independent non-linearity parameters is given by the number of distinct ways of applying $2(n-1)$ derivatives to $n$ factors of $N$ with at least one derivative acting on each factor of $N$.  This is equivalent to the number of ``free trees'' constructed from contracting $n$ vertices with $n-1$ edges, called $t_n$ in graph theory and combinatorial analysis \cite{Fry:1983cj,riordan2002introduction}.  We can write the set of non-linearity parameters for the $n$-point function as
\begin{equation}\label{npointparameters}
F_{\mathrm{NL},i}^{(n)}=\frac{\sum_{A_1,A_2,A_3,\ldots}N_{,A_1A_2\ldots}N_{,A_1A_3\ldots} \cdots N_{,A_2}N_{,A_3}}{\left(\sum_{K}N_{,K}^2\right)^{n-1}}\, ,
\end{equation}
where the index $i$ runs from $1$ to $t_n$ and labels the various ways of distributing $2(n-1)$ derivatives in the numerator.  We can make the identifications $F_{\mathrm{NL},1}^{(3)}=\frac{6}{5}f_{\mathrm{NL}}^{(4)}$, $F_{\mathrm{NL},1}^{(4)}=\tau_{\mathrm{NL}}$, and $F_{\mathrm{NL},2}^{(4)}=\frac{54}{25}g_{\mathrm{NL}}$.  There are 3 parameters for the five-point function, 6 for the six-point function, 11 for the seven-point function, 23 for the eight-point function, and so on \cite{Fry:1983cj,riordan2002introduction}.  We give explicit expressions for the five and six-point non-linearity parameters in appendix \ref{fiveandsix}.

The structure of the derivatives of $N$ follows a pattern which allows for an easy determination of the most important contributions to each of the non-linearity parameters.  Specifically, the $m$th derivative of $N$ will be of the form
\begin{equation}\label{mthderivative}
\frac{\partial^m N}{\partial \phi_*^m}=\sum_{k=0}^{m-1}\mathcal{O}\left(\varepsilon_*^{(2k-m)/2}\right) \, .
\end{equation}
Each of the non-linearity parameters describing the $n$-point function has $n$ factors of $N$ with a total of $2(n-1)$ derivatives in the numerator and $2(n-1)$ factors of $N_{,K}$ in the denominator.  This means that the denominator will be $\mathcal{O}\left(\varepsilon_*^{-(n-1)}\right)$, so only the terms which are $\mathcal{O}\left(\varepsilon_*^{-(n-1)}\right)$ or larger in the numerator will make a contribution which is not automatically slow roll suppressed.  Now if we have a product of $n$ factors of $N$ of the form $N^{(m_1)}N^{(m_2)}\cdots N^{(m_n)}$, where $N^{(m_l)}$ refers to the $m_l$th derivative of $N$ with respect to $\phi_*^I$, the term which is lowest order in slow roll parameters will be $\mathcal{O}\left(\varepsilon_*^{-\left(\sum_i m_i\right)/2}\right)$.  For a product of $n$ factors of $N$ with a total of $2(n-1)$ derivatives, the leading term is $\mathcal{O}\left(\varepsilon_*^{-(n-1)}\right)$, and so the leading term is the only one which is not automatically slow roll suppressed.

The leading term for all second and higher derivatives of $N$ contains a factor $\frac{\epsilon_c^{\phi}\epsilon_c^{\chi}}{\epsilon_c^2}$.  One can see this directly for the second and third derivatives by examining equations (\ref{secondderivativesh}-\ref{thirdderivativesh}) for the homogeneous potential and equations (\ref{secondderivativese}-\ref{thirdderivativese}) for the exponential potential.  It is straightforward to verify that derivatives of $\frac{\epsilon_c^{\phi}\epsilon_c^{\chi}}{\epsilon_c^2}$ with respect to $\phi_*$ and $\chi_*$ are proportional to a sum of slow roll parameters multiplied by the combination $\frac{\epsilon_c^{\phi}\epsilon_c^{\chi}}{\epsilon_c^2}$.  It is exactly this quantity which becomes exponentially suppressed as non-adiabatic fluctuations are damped away by passing through a phase of single field inflation.  As long as the terms which are multiplying $\frac{\epsilon_c^{\phi}\epsilon_c^{\chi}}{\epsilon_c^2}$ are not exponentially growing during the phase of single field inflation, we find that the leading term in all second and higher derivatives of $N$ becomes exponentially suppressed as non-adiabatic fluctuations are damped away.  As we discussed above, the various slow roll parameters do not evolve independently, and thus generally cannot grow exponentially without affecting the evolution of $\epsilon^{\phi}$ and $\epsilon^{\chi}$.  As a result, we can conclude that the leading contribution to all local form non-gaussian $n$-point functions becomes suppressed as non-adiabatic fluctuations are damped away during a phase of single field inflation, and thus all $F_{\mathrm{NL},i}^{(n)}$ will be slow roll suppressed.

There is a complication which we have so far overlooked, which is that there are other contributions to the non-linearity parameters which contain more than the minimum number of derivatives acting on the factors of $N$, see for example \cite{Cogollo:2008bi,Rodriguez:2008hy,Byrnes:2007tm,Byrnes:2008wi}.  These terms are generally known as loop corrections because they involve integrals over internal momenta.  These terms all contain second and higher derivatives of $N$, and are thus suppressed for the same reason as the leading terms discussed above.

\section{Conclusions}\label{conclusion}
We have extended the analysis of \cite{Meyers:2010rg} to include the calculation of the trispectrum.  We calculated $\tau_{\mathrm{NL}}$ and $g_{\mathrm{NL}}$ for inflationary potentials of the form $W(\phi,\chi)=\left[U(\phi)+V(\chi)\right]^{\gamma}$ and $W(\phi,\chi)=W_0\mathrm{Exp}\left[U(\phi)+V(\chi)\right]$ by using the $\delta N$ formalism.  We focused on the case in which there is a phase of effectively single field inflation which damps away non-adiabatic fluctuations before the end of inflation.  This ensures that all of the $n$-point statistics are conserved during the subsequent evolution of the universe, and it also guarantees that the power spectrum is purely adiabatic as indicated by observation.  Under these conditions, we find that $\tau_{\mathrm{NL}}$ and $g_{\mathrm{NL}}$ are always slow roll suppressed upon entering a purely adiabatic solution.  We also discussed the extension to higher point functions and argued that all of the non-linearity parameters $F_{\mathrm{NL},i}^{(n)}$ that describe the local form non-gaussianity for any $n$-point function will be slow roll suppressed after a phase of effectively single field inflation.

There are, of course, some serious limitations with our result. In particular, one might worry about different potentials, different (slow roll violating) trajectories and different approaches to adiabaticity. As discussed in \cite{Meyers:2010rg}, there are some arguments which suggest that the first two of these would not have significant effects on our results. In \cite{Meyers:2010rg} (and as recapped at the end of section \ref{themodel}) we show that small violations of slow roll do not affect the results and argue that larger ones would be inconsistent with the observed power spectrum. With regards to a different potential, it is unclear how such a change could alter the fundamentals of our analysis, though without a detailed study one cannot be certain of this.

It is, however, the particular approach to adiabaticity that is the most restrictive assumption we make. While our results illustrate the challenges associated with generating an adiabatic spectrum and local form non-gaussianities, there a number of scenarios where the fluctuations in the cosmological fluid enter an adiabatic mode without a phase of single field inflation. As discussed in the introduction these include (amongst others) curvaton models and a phase of thermal equilibrium. We hope to explore the evolution of the bispectrum, trispectrum and beyond in such scenarios in forthcoming work.



\begin{acknowledgments}
The authors would like to thank Eiichiro Komatsu for helpful discussions. This material is based upon work supported by the National Science Foundation under Grant No. PHY-0455649 and by the Texas Cosmology Center, which is supported by the College of Natural Sciences and the Department of Astronomy at the University of Texas at Austin and the McDonald Observatory.
\end{acknowledgments}

\appendix
\section{Details for Homogeneous and Exponential Potentials}\label{details}
\subsection*{Homogeneous Potential: $W(\phi,\chi)=\left[U(\phi)+V(\phi)\right]^{\gamma}$}
With $N$ and $C$ given by (\ref{Cpotential}) and (\ref{Nhomogeneous}):
\begin{align}
C&=-m_p^2\int_{\phi_0}^{\phi}\frac{1}{U'(\phi')}\textrm{d}\,\phi'+m_p^2\int_{\chi_0}^{\chi}\frac{1}{U'(\chi')}\,\textrm{d}\chi'\tag{\ref{Cpotential}}\, ,\\
N&=-\frac{1}{\gamma m_p^2}\int_*^c\frac{U(\phi)}{U'\left(\phi\right)}\textrm{d}\phi-\frac{1}{\gamma m_p^2}\int_*^c\frac{V(\chi)}{V'\left(\chi\right)}\textrm{d}\chi\, . \tag{\ref{Nhomogeneous}}
\end{align}
Varying $N$ then gives:
\begin{multline}\label{dNhomogeneous}
\textrm{d}N=\frac{1}{m_p^2\gamma}\left[\left(\frac{U}{U'}\right)_*-\frac{\partial\phi_c}{\partial\phi_*}\left(\frac{U}{U'}\right)_c-\frac{\partial\chi_c}{\partial\phi_*}\left(\frac{V}{V'}\right)_c\right]\textrm{d}\phi_*\\
+\frac{1}{m_p^2\gamma}\left[\left(\frac{V}{V'}\right)_*-\frac{\partial\phi_c}{\partial\chi_*}\left(\frac{U}{U'}\right)_c-\frac{\partial\chi_c}{\partial\chi_*}\left(\frac{V}{V'}\right)_c\right]\textrm{d}\chi_*\, .
\end{multline}
Note that in deriving the above we had to account for the dependence of $\phi_c$ and $\chi_c$ on both $\phi_*$ and $\chi_*$. We will also need:
\begin{align}\label{ctostar}
\textrm{d}\phi_c&=\frac{\textrm{d}\phi_c}{\textrm{d}C}\left(\frac{\partial C}{\partial\phi_*}\textrm{d}\phi_*+\frac{\partial C}{\partial\chi_*}\textrm{d}\chi_*\right)\, ,\nonumber\\
\textrm{d}\chi_c&=\frac{\textrm{d}\chi_c}{\textrm{d}C}\left(\frac{\partial C}{\partial\phi_*}\textrm{d}\phi_*+\frac{\partial C}{\partial\chi_*}\textrm{d}\chi_*\right)\, .
\end{align}
From (\ref{Cpotential}) we have:
\begin{align}\label{dCdphistar}
\frac{\partial C}{\partial\phi_*}&=-\frac{m_p^2}{U'_*}\, , & \frac{\partial C}{\partial\chi_*}&=\frac{m_p^2}{V'_*}\, ,
\end{align}
and
\begin{align}
\frac{\partial C}{\partial\phi_c}&=-\frac{m_p^2}{U'_c}\, , & \frac{\partial C}{\partial\chi_c}&=\frac{m_p^2}{V'_c}\, .
\end{align}
The time $t_c$ defines a surface of constant energy:
\begin{equation}
W\left(\phi_c,\chi_c\right)=\textrm{constant}\, .
\end{equation}
Differentiating with respect to $C$ then gives:
\begin{equation}
\frac{\textrm{d}\phi_c}{\textrm{d}C}\left.W_{\phi}\right|_c+\frac{\textrm{d}\chi_c}{\textrm{d}C}\left.W_{\chi}\right|_c=0\, .
\end{equation}
Using the above and $W_{\phi}/W_{\chi}=U'/V'$, we can differentiate the expression for $C$ in (\ref{Cpotential}) and obtain (after some manipulation) the following expressions:
\begin{gather}
\frac{\textrm{d}\phi_c}{\textrm{d}C}=-\frac{1}{m_p^2}\left[U'_c\left(\frac{1}{U^{'2}_c}+\frac{1}{V^{'2}_c}\right)\right]^{-1}=-\frac{1}{m_p^2}\frac{U'_cV^{'2}_c}{U^{'2}_c+V^{'2}_c}\, ,\nonumber\\
\frac{\textrm{d}\chi_c}{\textrm{d}C}=\frac{1}{m_p^2}\left[V'_c\left(\frac{1}{U^{'2}_c}+\frac{1}{V^{'2}_c}\right)\right]^{-1}=\frac{1}{m_p^2}\frac{U^{'2}_cV'_c}{U^{'2}_c+V^{'2}_c}\, .\label{dphidCc}
\end{gather}
Substituting (\ref{dCdphistar}) and (\ref{dphidCc}) into (\ref{ctostar}) allows us to read off the following:
\begin{align}\label{dcdstar}
\frac{\partial\phi_c}{\partial\phi_*}&=\frac{V^{'2}_c}{U^{'2}_c+V^{'2}_c}\frac{U_c'}{U_*'}\, , & \frac{\partial\phi_c}{\partial\chi_*}&=-\frac{V^{'2}_c}{U^{'2}_c+V^{'2}_c}\frac{U_c'}{V_*'}\, ,\nonumber\\
\frac{\partial\chi_c}{\partial\phi_*}&=-\frac{U^{'2}_c}{U^{'2}_c+V^{'2}_c}\frac{V_c'}{U_*'}\, , & \frac{\partial\chi_c}{\partial\chi_*}&=\frac{U^{'2}_c}{U^{'2}_c+V^{'2}_c}\frac{V_c'}{V_*'}\, .
\end{align}
The derivatives of $N$ are then:
\begin{align}
\frac{\partial N}{\partial\phi_*}&=\frac{1}{m_p}\frac{x_h}{\sqrt{2\epsilon_*^{\phi}}}\, , & \frac{\partial N}{\partial\chi_*}&=\frac{1}{m_p}\frac{y_h}{\sqrt{2\epsilon_*^{\chi}}}\, .
\end{align}
We have used the slow roll parameters defined in (\ref{slowroll}); while $x_h$ and $y_h$ are defined in (\ref{xy}):
\begin{align}
x_h&\equiv\frac{1}{U_*+V_*}\left(U_*+\frac{V_c\epsilon_c^{\phi}-U_c\epsilon_c^{\chi}}{\epsilon_c}\right)\, ,\nonumber\\
y_h&\equiv\frac{1}{U_*+V_*}\left(V_*-\frac{V_c\epsilon_c^{\phi}-U_c\epsilon_c^{\chi}}{\epsilon_c}\right)\, . \tag{\ref{xy}}
\end{align}
In a similar vein, we can find the second derivatives:
\begin{align}
\begin{split}
\frac{\partial^2N}{\partial\phi_*^2}&=\frac{1}{\gamma m_p^2}\Bigg[1-x_h\left(\frac{\gamma\eta_*^{\phi}}{2\epsilon_*^{\phi}}-\gamma+1\right)\\
&\qquad\qquad+\frac{\left(U_c+V_c\right)^2}{\left(U_*+V_*\right)^2}\frac{1}{\epsilon_*^{\phi}}\frac{\epsilon_c^{\phi}\epsilon_c^{\chi}}{\epsilon_c}\left(\frac{\gamma\eta^{ss}_c}{\epsilon_c}-1\right)\Bigg]\, ,\nonumber
\end{split}\\
\begin{split}
\frac{\partial^2N}{\partial\chi_*^2}&=\frac{1}{\gamma m_p^2}\Bigg[1-y_h\left(\frac{\gamma\eta_*^{\chi}}{2\epsilon_*^{\chi}}-\gamma+1\right)\\
&\qquad\qquad+\frac{\left(U_c+V_c\right)^2}{\left(U_*+V_*\right)^2}\frac{1}{\epsilon_*^{\chi}}\frac{\epsilon_c^{\phi}\epsilon_c^{\chi}}{\epsilon_c}\left(\frac{\gamma\eta^{ss}_c}{\epsilon_c}-1\right)\Bigg]\, ,\nonumber
\end{split}\\
\frac{\partial^2N}{\partial\phi_*\chi_*}&=\frac{1}{\gamma m_p^2}\left[-\frac{\left(U_c+V_c\right)^2}{\left(U_*+V_*\right)^2}\frac{1}{\sqrt{\epsilon_*^{\phi}\epsilon_*^{\chi}}}\frac{\epsilon_c^{\phi}\epsilon_c^{\chi}}{\epsilon_c}\left(\frac{\gamma\eta^{ss}_c}{\epsilon_c}-1\right)\right]\, \label{secondderivativesh},
\end{align}
and third derivatives:
\begin{widetext}
\begin{multline}
\frac{\partial^3N}{\partial\phi_*^3}=\frac{1}{\gamma m_p^2}\left[-\frac{1}{m_p}\frac{\eta_*^{\phi}}{\sqrt{2\epsilon_*^{\phi}}}+\frac{\gamma-1}{\gamma m_p}\sqrt{2\epsilon_*^{\phi}}-\frac{x_h}{m_p}\left(\frac{\gamma}{2\sqrt{2}}\frac{\left.\xi_*^{\phi}\right.^2}{\left.\epsilon_*^{\phi}\right.^{3/2}}-\frac{\gamma}{\sqrt{2}}\frac{\left.\eta_*^{\phi}\right.^2}{\left.\epsilon_*^{\phi}\right.^{3/2}}+\frac{\gamma-1}{m_p}\frac{\eta_*^{\phi}}{\sqrt{2\epsilon_*^{\phi}}}+\frac{\gamma-1}{\gamma m_p}\sqrt{2\epsilon_*^{\phi}}\right)\right.\\
-3\left(\frac{1}{m_p}\frac{\eta_*^{\phi}}{\sqrt{2\epsilon_*^{\phi}}}+\frac{\gamma-1}{\gamma m_p}\sqrt{2\epsilon_*^{\phi}}\right)\frac{\left(U_c+V_c\right)^2}{\left(U_*+V_*\right)^2}\frac{\epsilon_c^{\phi}\epsilon_c^{\chi}}{\epsilon_c}\frac{1}{\epsilon_*^{\phi}}\left(\frac{\gamma\eta_c^{ss}}{\epsilon_c}-1\right)\\
+\frac{\gamma}{m_p}\frac{1}{\sqrt{2}\left.\epsilon_*^{\phi}\right.^{3/2}}\frac{\left(U_c+V_c\right)^3}{\left(U_*+V_*\right)^3}\frac{\epsilon_c^{\phi}\epsilon_c^{\chi}}{\epsilon_c^4}\left\{\frac{2}{\gamma}\epsilon\left(\left.\epsilon^{\phi}\right.^2\eta^{\chi}-\left.\epsilon^{\chi}\right.^2\eta^{\phi}\right)+\frac{4\left(\gamma-1\right)}{\gamma^2}\epsilon^{\phi}\epsilon^{\chi}\left(\left.\epsilon^{\chi}\right.^2-\left.\epsilon^{\phi}\right.^2\right)\right.\\
+2\left.\epsilon^{\chi}\right.^2\left.\eta^{\phi}\right.^2-2\left.\epsilon^{\phi}\right.^2\left.\eta^{\chi}\right.^2+\frac{14\left(\gamma-1\right)}{\gamma^2}\epsilon^{\phi}\epsilon^{\chi}\left(\epsilon^{\phi}\eta^{\chi}-\epsilon^{\chi}\eta^{\phi}\right)+\left.\epsilon^{\chi}\right.^2\left.\xi^{\phi}\right.^2-\left.\epsilon^{\phi}\right.^2\left.\xi^{\chi}\right.^2\\
\left.6\epsilon^{\phi}\epsilon^{\chi}\eta^{ss}\left(\eta^{\chi}-\eta^{\phi}+\frac{2\left(\gamma-1\right)}{\gamma}\left(\epsilon^{\phi}-\epsilon^{\chi}\right)\right)\right\}_c\Bigg]\, ,
\end{multline}
\begin{multline}
\frac{\partial^3N}{\partial\chi_*^3}=\frac{1}{\gamma m_p^2}\left[-\frac{1}{m_p}\frac{\eta_*^{\chi}}{\sqrt{2\epsilon_*^{\chi}}}+\frac{\gamma-1}{\gamma m_p}\sqrt{2\epsilon_*^{\chi}}-\frac{y_h}{m_p}\left(\frac{\gamma}{2\sqrt{2}}\frac{\left.\xi_*^{\chi}\right.^2}{\left.\epsilon_*^{\chi}\right.^{3/2}}-\frac{\gamma}{\sqrt{2}}\frac{\left.\eta_*^{\chi}\right.^2}{\left.\epsilon_*^{\chi}\right.^{3/2}}+\frac{\gamma-1}{m_p}\frac{\eta_*^{\chi}}{\sqrt{2\epsilon_*^{\chi}}}+\frac{\gamma-1}{\gamma m_p}\sqrt{2\epsilon_*^{\chi}}\right)\right.\\
-3\left(\frac{1}{m_p}\frac{\eta_*^{\chi}}{\sqrt{2\epsilon_*^{\chi}}}+\frac{\gamma-1}{\gamma m_p}\sqrt{2\epsilon_*^{\chi}}\right)\frac{\left(U_c+V_c\right)^2}{\left(U_*+V_*\right)^2}\frac{\epsilon_c^{\chi}\epsilon_c^{\phi}}{\epsilon_c}\frac{1}{\epsilon_*^{\chi}}\left(\frac{\gamma\eta_c^{ss}}{\epsilon_c}-1\right)\\
+\frac{\gamma}{m_p}\frac{1}{\sqrt{2}\left.\epsilon_*^{\chi}\right.^{3/2}}\frac{\left(U_c+V_c\right)^3}{\left(U_*+V_*\right)^3}\frac{\epsilon_c^{\chi}\epsilon_c^{\phi}}{\epsilon_c^4}\left\{\frac{2}{\gamma}\epsilon\left(\left.\epsilon^{\chi}\right.^2\eta^{\phi}-\left.\epsilon^{\phi}\right.^2\eta^{\chi}\right)+\frac{4\left(\gamma-1\right)}{\gamma^2}\epsilon^{\chi}\epsilon^{\phi}\left(\left.\epsilon^{\phi}\right.^2-\left.\epsilon^{\chi}\right.^2\right)\right.\\
+2\left.\epsilon^{\phi}\right.^2\left.\eta^{\chi}\right.^2-2\left.\epsilon^{\chi}\right.^2\left.\eta^{\phi}\right.^2+\frac{14\left(\gamma-1\right)}{\gamma^2}\epsilon^{\chi}\epsilon^{\phi}\left(\epsilon^{\chi}\eta^{\phi}-\epsilon^{\phi}\eta^{\chi}\right)+\left.\epsilon^{\phi}\right.^2\left.\xi^{\chi}\right.^2-\left.\epsilon^{\chi}\right.^2\left.\xi^{\phi}\right.^2\\
\left.6\epsilon^{\chi}\epsilon^{\phi}\eta^{ss}\left(\eta^{\phi}-\eta^{\chi}+\frac{2\left(\gamma-1\right)}{\gamma}\left(\epsilon^{\chi}-\epsilon^{\phi}\right)\right)\right\}_c\Bigg]\, ,
\end{multline}
\begin{multline}
\frac{\partial^3N}{\partial\phi_*^2\partial\chi_*}=\frac{1}{\gamma m_p^2}\left[\left(\frac{1}{m_p}\frac{\eta_*^{\phi}}{\sqrt{2\epsilon_*^{\phi}}}+\frac{\gamma-1}{\gamma m_p}\sqrt{2\epsilon_*^{\phi}}\right)\frac{\left(U_c+V_c\right)^2}{\left(U_*+V_*\right)^2}\frac{\epsilon_c^{\phi}\epsilon_c^{\chi}}{\epsilon_c}\frac{1}{\sqrt{\epsilon_*^{\phi}\epsilon_*^{\chi}}}\left(\frac{\gamma\eta_c^{ss}}{\epsilon_c}-1\right)\right.\\
-\frac{\gamma}{m_p}\frac{1}{\sqrt{2\epsilon_*^{\chi}}\epsilon_*^{\phi}}\frac{\left(U_c+V_c\right)^3}{\left(U_*+V_*\right)^3}\frac{\epsilon_c^{\phi}\epsilon_c^{\chi}}{\epsilon_c^4}\left\{\frac{2}{\gamma}\epsilon\left(\left.\epsilon^{\phi}\right.^2\eta^{\chi}-\left.\epsilon^{\chi}\right.^2\eta^{\phi}\right)+\frac{4\left(\gamma-1\right)}{\gamma^2}\epsilon^{\phi}\epsilon^{\chi}\left(\left.\epsilon^{\chi}\right.^2-\left.\epsilon^{\phi}\right.^2\right)\right.\\
+2\left.\epsilon^{\chi}\right.^2\left.\eta^{\phi}\right.^2-2\left.\epsilon^{\phi}\right.^2\left.\eta^{\chi}\right.^2+\frac{14\left(\gamma-1\right)}{\gamma^2}\epsilon^{\phi}\epsilon^{\chi}\left(\epsilon^{\phi}\eta^{\chi}-\epsilon^{\chi}\eta^{\phi}\right)+\left.\epsilon^{\chi}\right.^2\left.\xi^{\phi}\right.^2-\left.\epsilon^{\phi}\right.^2\left.\xi^{\chi}\right.^2\\
\left.6\epsilon^{\phi}\epsilon^{\chi}\eta^{ss}\left(\eta^{\chi}-\eta^{\phi}+\frac{2\left(\gamma-1\right)}{\gamma}\left(\epsilon^{\phi}-\epsilon^{\chi}\right)\right)\right\}_c\Bigg]\, ,
\end{multline}
\begin{multline}
\frac{\partial^3N}{\partial\chi_*^2\partial\phi_*}=\frac{1}{\gamma m_p^2}\left[\left(\frac{1}{m_p}\frac{\eta_*^{\chi}}{\sqrt{2\epsilon_*^{\chi}}}+\frac{\gamma-1}{\gamma m_p}\sqrt{2\epsilon_*^{\chi}}\right)\frac{\left(U_c+V_c\right)^2}{\left(U_*+V_*\right)^2}\frac{\epsilon_c^{\chi}\epsilon_c^{\phi}}{\epsilon_c}\frac{1}{\sqrt{\epsilon_*^{\chi}\epsilon_*^{\phi}}}\left(\frac{\gamma\eta_c^{ss}}{\epsilon_c}-1\right)\right.\\
-\frac{\gamma}{m_p}\frac{1}{\sqrt{2\epsilon_*^{\phi}}\epsilon_*^{\chi}}\frac{\left(U_c+V_c\right)^3}{\left(U_*+V_*\right)^3}\frac{\epsilon_c^{\chi}\epsilon_c^{\phi}}{\epsilon_c^4}\left\{\frac{2}{\gamma}\epsilon\left(\left.\epsilon^{\chi}\right.^2\eta^{\phi}-\left.\epsilon^{\phi}\right.^2\eta^{\chi}\right)+\frac{4\left(\gamma-1\right)}{\gamma^2}\epsilon^{\chi}\epsilon^{\phi}\left(\left.\epsilon^{\phi}\right.^2-\left.\epsilon^{\chi}\right.^2\right)\right.\\
+2\left.\epsilon^{\phi}\right.^2\left.\eta^{\chi}\right.^2-2\left.\epsilon^{\chi}\right.^2\left.\eta^{\phi}\right.^2+\frac{14\left(\gamma-1\right)}{\gamma^2}\epsilon^{\chi}\epsilon^{\phi}\left(\epsilon^{\chi}\eta^{\phi}-\epsilon^{\phi}\eta^{\chi}\right)+\left.\epsilon^{\phi}\right.^2\left.\xi^{\chi}\right.^2-\left.\epsilon^{\chi}\right.^2\left.\xi^{\phi}\right.^2\\
\left.6\epsilon^{\chi}\epsilon^{\phi}\eta^{ss}\left(\eta^{\phi}-\eta^{\chi}+\frac{2\left(\gamma-1\right)}{\gamma}\left(\epsilon^{\chi}-\epsilon^{\phi}\right)\right)\right\}_c\Bigg]\, \label{thirdderivativesh},
\end{multline}
where $\xi$ is defined by (\ref{xi}):
\begin{align}
\left.\xi^{\phi}\right.^2&\equiv m_p^4\frac{W_{,\phi}W_{,\phi\phi\phi}}{W^2}=m_p^4\frac{\gamma^2\left(U+V\right)^2U'U'''+3\gamma^2\left(\gamma-1\right)\left(U+V\right)U'^2U''+\gamma^2\left(\gamma-1\right)\left(\gamma-2\right)U'^4}{\left(U+V\right)^4}\nonumber\\
\left.\xi^{\chi}\right.^2&\equiv m_p^4\frac{W_{,\chi}W_{,\chi\chi\chi}}{W^2}=m_p^4\frac{\gamma^2\left(U+V\right)^2V'V'''+3\gamma^2\left(\gamma-1\right)\left(U+V\right)V'^2V''+\gamma^2\left(\gamma-1\right)\left(\gamma-2\right)V'^4}{\left(U+V\right)^4}\, ,\tag{\ref{xi}}
\end{align}
and $\eta^{ss}$ is given in (\ref{etass2}):
\begin{align}
\eta^{ss}&=\frac{\epsilon^{\chi}\eta^{\phi}-4\frac{(\gamma-1)}{\gamma}\epsilon^{\phi}\epsilon^{\chi}+\epsilon^{\phi}\eta^{\chi}}{\epsilon}\, .\tag{\ref{etass2}}
\end{align}
From these it is straightforward to use the relevant $\delta N$ equations to obtain expressions for $f_{\mathrm{NL}}$, $\tau_{\mathrm{NL}}$ and $g_{\mathrm{NL}}$:
\begin{multline}
\frac{6}{5}f_{\mathrm{NL}}^{(4)}= \frac{2}{\gamma}\left(\frac{\frac{x_h^2}{\epsilon^{\phi}_*}\left[1-\left(\frac{\gamma\eta^{\phi}_*}{2\epsilon_*^{\phi}}-\gamma+1\right)x_h\right]+\frac{y_h^2}{\epsilon^{\chi}_*}\left[1-\left(\frac{\gamma\eta^{\chi}_*}{2\epsilon_*^{\chi}}-\gamma+1\right)y_h\right]}{\left(\frac{x_h^2}{\epsilon_*^{\phi}}+\frac{y_h^2}{\epsilon_*^{\chi}}\right)^2}\right)\\
+\frac{2}{\gamma}\left(\frac{\frac{\left(U_c+V_c\right)^2}{\left(U_*+V_*\right)^2}\left(\frac{x_h}{\epsilon_*^{\phi}}-\frac{y_h}{\epsilon_*^{\chi}}\right)^2\frac{\epsilon_c^{\phi}\epsilon_c^{\chi}}{\epsilon_c}\left(\frac{\gamma\eta_c^{ss}}{\epsilon_c}-1\right)}{\left(\frac{x_h^2}{\epsilon_*^{\phi}}+\frac{y_h^2}{\epsilon_*^{\chi}}\right)^2}\right)\, ,
\end{multline}
\begin{multline}
\tau_{\mathrm{NL}}=\frac{4}{\gamma^2}\left(\frac{\left[1-x_h\left(\frac{\gamma\eta_*^{\phi}}{2\epsilon_*^{\phi}}-\gamma+1\right)\right]^2\frac{x_h^2}{\epsilon_*^{\phi}}+\left[1-y_h\left(\frac{\gamma\eta_*^{\chi}}{2\epsilon_*^{\chi}}-\gamma+1\right)\right]^2\frac{y_h^2}{\epsilon_*^{\chi}}}{\left(\frac{x_h^2}{\epsilon_*^{\phi}}+\frac{y_h^2}{\epsilon_*^{\chi}}\right)^3}\right)\\
+\frac{8}{\gamma^2}\left(\frac{\left[1-x_h\left(\frac{\gamma\eta_*^{\phi}}{2\epsilon_*^{\phi}}-\gamma+1\right)\right]^2\frac{x_h}{\epsilon_*^{\phi}}\left(\frac{x_h}{\epsilon_*^{\phi}}-\frac{y_h}{\epsilon_*^{\chi}}\right)+\left[1-y_h\left(\frac{\gamma\eta_*^{\chi}}{2\epsilon_*^{\chi}}-\gamma+1\right)\right]^2\frac{y_h}{\epsilon_*^{\chi}}\left(\frac{y_h}{\epsilon_*^{\chi}}-\frac{x_h}{\epsilon_*^{\phi}}\right)}{\left(\frac{x_h^2}{\epsilon_*^{\phi}}+\frac{y_h^2}{\epsilon_*^{\chi}}\right)^3}\right)\left[\frac{\left(U_c+V_c\right)^2}{\left(U_*+V_*\right)^2}\frac{\epsilon_c^{\phi}\epsilon_c^{\chi}}{\epsilon_c}\left(\frac{\gamma\eta^{ss}_c}{\epsilon_c}-1\right)\right]\\
+\frac{4}{\gamma^2}\left(\frac{\left[\frac{\left(U_c+V_c\right)^2}{\left(U_*+V_*\right)^2}\frac{\epsilon_c^{\phi}\epsilon_c^{\chi}}{\epsilon_c}\left(\frac{\gamma\eta^{ss}_c}{\epsilon_c}-1\right)\right]^2\left(\frac{x_h}{\epsilon_*^{\phi}}-\frac{y_h}{\epsilon_*^{\chi}}\right)^2\left(\frac{1}{\epsilon_*^{\phi}}+\frac{1}{\epsilon_*^{\chi}}\right)}{\left(\frac{x_h^2}{\epsilon_*^{\phi}}+\frac{y_h^2}{\epsilon_*^{\chi}}\right)^3}\right)\, ,
\end{multline}
and
\begin{multline}
g_{\mathrm{NL}}=\frac{50}{27}\frac{1}{\gamma}\frac{1}{\left(\frac{x_h^2}{\epsilon_*^{\phi}}+\frac{y_h^2}{\epsilon_*^{\chi}}\right)^3}\\
\times\left\{\frac{x_h^3}{\epsilon_*^{\phi}}\left[-\frac{\eta_*^{\phi}}{2\epsilon_*^{\phi}}+\frac{\gamma-1}{\gamma}-x_h\left(\frac{\gamma\left.\xi_*^{\phi}\right.^2}{4\left.\epsilon_*^{\phi}\right.^2}-\frac{\gamma\left.\eta_*^{\phi}\right.^2}{2\left.\epsilon_*^{\phi}\right.^2}+\frac{\left(\gamma-1\right)\eta_*^{\phi}}{2\epsilon_*^{\phi}}+\frac{\gamma-1}{\gamma}\right)\right]\right.\\
\frac{y_h^3}{\epsilon_*^{\chi}}\left[-\frac{\eta_*^{\chi}}{2\epsilon_*^{\chi}}+\frac{\gamma-1}{\gamma}-y_h\left(\frac{\gamma\left.\xi_*^{\chi}\right.^2}{4\left.\epsilon_*^{\chi}\right.^2}-\frac{\gamma\left.\eta_*^{\chi}\right.^2}{2\left.\epsilon_*^{\chi}\right.^2}+\frac{\left(\gamma-1\right)\eta_*^{\chi}}{2\epsilon_*^{\chi}}+\frac{\gamma-1}{\gamma}\right)\right]\\
+3\left[\left(\frac{x_h^2y_h^{}}{\epsilon_*^{\phi}\epsilon_*^{\chi}}-\frac{x_h^3}{\left.\epsilon_*^{\phi}\right.^2}\right)\left(\frac{\eta_*^{\phi}}{2\epsilon_*^{\phi}}-\frac{\gamma-1}{\gamma}\right)+\left(\frac{y_h^2x_h^{}}{\epsilon_*^{\phi}\epsilon_*^{\chi}}-\frac{y_h^3}{\left.\epsilon_*^{\chi}\right.^2}\right)\left(\frac{\eta_*^{\chi}}{2\epsilon_*^{\chi}}-\frac{\gamma-1}{\gamma}\right)\right]\frac{\left(U_c+V_c\right)^2}{\left(U_*+V_*\right)^2}\frac{\epsilon_c^{\phi}\epsilon_c^{\chi}}{\epsilon_c}\left(\frac{\gamma\eta^{ss}_c}{\epsilon_c}-1\right)\\
+\left(\frac{x_h}{\epsilon_*^{\phi}}-\frac{y_h}{\epsilon_*^{\chi}}\right)^3\frac{\left(U_c+V_c\right)^3}{\left(U_*+V_*\right)^3}\frac{\epsilon_c^{\phi}\epsilon_c^{\chi}}{\epsilon_c^4}\\
\times\left[-\epsilon\left.\epsilon^{\chi}\right.^2\eta^{\phi}+\epsilon\left.\epsilon^{\phi}\right.^2\eta^{\chi}+\frac{2\left(\gamma-1\right)}{\gamma}\epsilon^{\phi}\epsilon^{\chi}\left(\epsilon^{\chi}-\epsilon^{\phi}\right)+\left.\epsilon^{\chi}\right.^2\eta^{\phi}\left(\gamma\eta^{\phi}-\frac{7\left(\gamma-1\right)}{\gamma}\epsilon^{\phi}\right)-\left.\epsilon^{\phi}\right.^2\eta^{\chi}\left(\gamma\eta^{\chi}-\frac{7\left(\gamma-1\right)}{\gamma}\epsilon^{\chi}\right)\right.\\
\left.\left.+\frac{\gamma}{2}\left(\left.\epsilon^{\chi}\right.^2\left.\xi^{\phi}\right.^2-\left.\epsilon^{\phi}\right.^2\left.\xi^{\chi}\right.^2\right)+3\gamma\epsilon^{\phi}\epsilon^{\chi}\eta^{ss}\left(\eta^{\chi}-\frac{2\left(\gamma-1\right)}{\gamma}\epsilon^{\chi}-\eta^{\phi}+\frac{2\left(\gamma-1\right)}{\gamma}\epsilon^{\phi}\right)\right]_c\right\}\, .
\end{multline}
Eliminating the slow roll suppressed terms gives the expressions in (\ref{fnlhomogeneous}), (\ref{taunlhomogeneous}) and (\ref{gnlhomogeneous}). 
\end{widetext}

\subsection*{Exponential Potential: $W(\phi,\chi)=W_0\textrm{Exp}\left[U(\phi)+V(\chi)\right]$}
With $N$ from (\ref{Nexpo}) and $C$ again given by (\ref{Cpotential}),
\begin{align}
C&=-m_p^2\int_{\phi_0}^{\phi}\frac{1}{U'(\phi')}\textrm{d}\,\phi'+m_p^2\int_{\chi_0}^{\chi}\frac{1}{U'(\chi')}\,\textrm{d}\chi'\tag{\ref{Cpotential}}\\
N&=-\frac{1}{2m_p^2}\int_*^c\frac{1}{U'\left(\phi\right)}\textrm{d}\phi
-\frac{1}{2m_p^2}\int_*^c\frac{1}{V'\left(\chi\right)}\textrm{d}\chi\, ,\tag{\ref{Nexpo}}
\end{align}
the variation gives:
\begin{multline}\label{dNexpo}
\textrm{d}N=\frac{1}{2m_p^2}\left[\left(\frac{1}{U'}\right)_*-\frac{\partial\phi_c}{\partial\phi_*}\left(\frac{1}{U'}\right)_c-\frac{\partial\chi_c}{\partial\phi_*}\left(\frac{1}{V'}\right)_c\right]\textrm{d}\phi_*\\
+\frac{1}{2m_p^2}\left[\left(\frac{1}{V'}\right)_*-\frac{\partial\phi_c}{\partial\chi_*}\left(\frac{1}{U'}\right)_c-\frac{\partial\chi_c}{\partial\chi_*}\left(\frac{1}{V'}\right)_c\right]\textrm{d}\chi_*\, .
\end{multline}
The analysis begins in the same fashion as in the previous section, with the expressions for the derivatives of $\phi_c$ and $\chi_c$ with respect to $\phi_*$ and $\chi_*$, given in (\ref{dcdstar}). From this and (\ref{dNexpo}) we obtain the following expressions for the first derivatives of $N$:
\begin{align}
\frac{\partial N}{\partial\phi_*}&=\frac{1}{2m_p}\frac{x_e}{\sqrt{2\epsilon_*^{\phi}}}\, , & \frac{\partial N}{\partial\chi_*}&=\frac{1}{2m_p}\frac{y_e}{\sqrt{2\epsilon_*^{\chi}}}\, .
\end{align}
The slow roll parameters are defined in (\ref{slowroll}). $x_e$ and $y_e$ are defined in (\ref{xye}) as:
\begin{align}
x_e&\equiv \frac{2\epsilon_c^{\phi}}{\epsilon_c} & y_e\equiv \frac{2\epsilon_c^{\chi}}{\epsilon_c}\, . \tag{\ref{xye}}
\end{align}
The second derivatives are then given by:
\begin{align}
\frac{\partial^2N}{\partial\phi_*^2}&=\frac{1}{2m_p^2}\frac{1}{2\epsilon_*^{\phi}}\left[-\left(\eta_*^{\phi}-2\epsilon_*^{\phi}\right)x_e+\frac{4\epsilon_c^{\phi}\epsilon_c^{\chi}}{\epsilon_c^2}\eta^{ss}_c\right]\, ,\nonumber\\
\frac{\partial^2N}{\partial\chi_*^2}&=\frac{1}{2m_p^2}\frac{1}{2\epsilon_*^{\chi}}\left[-\left(\eta_*^{\chi}-2\epsilon_*^{\chi}\right)y_e+\frac{4\epsilon_c^{\phi}\epsilon_c^{\chi}}{\epsilon_c^2}\eta^{ss}_c\right]\, ,\nonumber\\
\frac{\partial^2N}{\partial\phi_*\chi_*}&=\frac{1}{2m_p^2}\frac{1}{\sqrt{\epsilon_*^{\phi}\epsilon_*^{\chi}}}\frac{2\epsilon_c^{\phi}\epsilon_c^{\chi}}{\epsilon_c^2}\eta^{ss}_c\, \label{secondderivativese}.
\end{align}
And the third derivatives are:
\begin{widetext}
\begin{multline}
\frac{\partial^3N}{\partial\phi_*^3}=\frac{1}{m_p^3\sqrt{2\left.\epsilon_*^{\chi}\right.^3}}\Bigg[-\frac{\epsilon_c^{\phi}}{2\epsilon_c}\left(2\left(\epsilon_*^{\phi}-\eta_*^{\phi}\right)\eta_*^{\phi}+\left.\xi_*^\phi\right.^2\right)+3\left(2\epsilon_*^{\phi}-\eta_*^{\phi}\right)\frac{\epsilon_c^{\phi}\epsilon_c^{\chi}}{\epsilon_c^2}\eta_c^{ss}\\
+\frac{\epsilon_c^{\phi}\epsilon_c^{\chi}}{\epsilon_c^5}\Bigg(\left.\epsilon_c^{\chi}\right.^3\left(2\left.\eta_c^{\phi}\right.^2+\left.\xi_c^{\phi}\right.^2\right)+\epsilon_c^{\phi}\left.\epsilon_c^{\chi}\right.^2\left(-26\epsilon_c^{\chi}\eta_c^{\phi}-4\left.\eta_c^{\phi}\right.^2+6\eta_c^{\phi}\eta_c^{\chi}+\left.\xi_c^{\phi}\right.^2\right)\\
+\left.\epsilon_c^{\phi}\right.^2\epsilon_c^{\chi}\left(48\left.\epsilon_c^{\chi}\right.^2+22\epsilon_c^{\chi}\left(\eta_c^{\phi}-\eta_c^{\chi}\right)-6\eta_c^{\phi}\eta_c^{\chi}+4\left.\eta_c^{\chi}\right.^2-\left.\xi_c^{\chi}\right.^2\right)-\left.\epsilon_c^{\phi}\right.^3\left(48\left.\epsilon_c^{\chi}\right.^2-26\epsilon_c^{\chi}\eta_c^{\chi}+2\left.\eta_c^{\chi}\right.^2+\left.\xi_c^{\chi}\right.^2\right)\Bigg)\Bigg]\, ,
\end{multline}
\begin{multline}
\frac{\partial^3N}{\partial\chi_*^3}=\frac{1}{m_p^3\sqrt{2\left.\epsilon_*^{\phi}\right.^3}}\Bigg[-\frac{\epsilon_c^{\chi}}{2\epsilon_c}\left(2\left(\epsilon_*^{\chi}-\eta_*^{\chi}\right)\eta_*^{\chi}+\left.\xi_*^\chi\right.^2\right)+3\left(2\epsilon_*^{\chi}-\eta_*^{\chi}\right)\frac{\epsilon_c^{\chi}\epsilon_c^{\phi}}{\epsilon_c^2}\eta_c^{ss}\\
+\frac{\epsilon_c^{\chi}\epsilon_c^{\phi}}{\epsilon_c^5}\Bigg(\left.\epsilon_c^{\phi}\right.^3\left(2\left.\eta_c^{\chi}\right.^2+\left.\xi_c^{\chi}\right.^2\right)+\epsilon_c^{\chi}\left.\epsilon_c^{\phi}\right.^2\left(-26\epsilon_c^{\phi}\eta_c^{\chi}-4\left.\eta_c^{\chi}\right.^2+6\eta_c^{\chi}\eta_c^{\phi}+\left.\xi_c^{\chi}\right.^2\right)\\
+\left.\epsilon_c^{\chi}\right.^2\epsilon_c^{\phi}\left(48\left.\epsilon_c^{\phi}\right.^2+22\epsilon_c^{\phi}\left(\eta_c^{\chi}-\eta_c^{\phi}\right)-6\eta_c^{\chi}\eta_c^{\phi}+4\left.\eta_c^{\phi}\right.^2-\left.\xi_c^{\phi}\right.^2\right)-\left.\epsilon_c^{\chi}\right.^3\left(48\left.\epsilon_c^{\phi}\right.^2-26\epsilon_c^{\phi}\eta_c^{\phi}+2\left.\eta_c^{\phi}\right.^2+\left.\xi_c^{\phi}\right.^2\right)\Bigg)\Bigg]\, ,
\end{multline}
\begin{multline}
\frac{\partial^3N}{\partial\phi_*^2\chi_*}=\frac{1}{m_p^3\epsilon_*^{\phi}\sqrt{2\epsilon_*^{\chi}}}\Bigg[-\left(2\epsilon_*^{\phi}-\eta_*^{\phi}\right)\frac{\epsilon_c^{\phi}\epsilon_c^{\chi}}{\epsilon_c^2}\eta_c^{ss}+\frac{\epsilon_c^{\phi}\epsilon_c^{\chi}}{\epsilon_c^5}\Bigg(\left.\epsilon_c^{\phi}\right.^3\left(2\left.\eta_c^{\chi}\right.^2+\left.\xi_c^{\chi}\right.^2\right)+\epsilon_c^{\chi}\left.\epsilon_c^{\phi}\right.^2\left(-26\epsilon_c^{\phi}\eta_c^{\chi}-4\left.\eta_c^{\chi}\right.^2+6\eta_c^{\chi}\eta_c^{\phi}+\left.\xi_c^{\chi}\right.^2\right)\\
+\left.\epsilon_c^{\chi}\right.^2\epsilon_c^{\phi}\left(48\left.\epsilon_c^{\phi}\right.^2+22\epsilon_c^{\phi}\left(\eta_c^{\chi}-\eta_c^{\phi}\right)-6\eta_c^{\chi}\eta_c^{\phi}+4\left.\eta_c^{\phi}\right.^2-\left.\xi_c^{\phi}\right.^2\right)-\left.\epsilon_c^{\chi}\right.^3\left(48\left.\epsilon_c^{\phi}\right.^2-26\epsilon_c^{\phi}\eta_c^{\phi}+2\left.\eta_c^{\phi}\right.^2+\left.\xi_c^{\phi}\right.^2\right)\Bigg)\Bigg]\, ,
\end{multline}
\begin{multline}
\frac{\partial^3N}{\partial\chi_*^2\phi_*}=\frac{1}{m_p^3\epsilon_*^{\chi}\sqrt{2\epsilon_*^{\phi}}}\Bigg[-\left(2\epsilon_*^{\chi}-\eta_*^{\chi}\right)\frac{\epsilon_c^{\chi}\epsilon_c^{\phi}}{\epsilon_c^2}\eta_c^{ss}+\frac{\epsilon_c^{\chi}\epsilon_c^{\phi}}{\epsilon_c^5}\Bigg(\left.\epsilon_c^{\chi}\right.^3\left(2\left.\eta_c^{\phi}\right.^2+\left.\xi_c^{\phi}\right.^2\right)+\epsilon_c^{\phi}\left.\epsilon_c^{\chi}\right.^2\left(-26\epsilon_c^{\chi}\eta_c^{\phi}-4\left.\eta_c^{\phi}\right.^2+6\eta_c^{\phi}\eta_c^{\chi}+\left.\xi_c^{\phi}\right.^2\right)\\
+\left.\epsilon_c^{\phi}\right.^2\epsilon_c^{\chi}\left(48\left.\epsilon_c^{\chi}\right.^2+22\epsilon_c^{\chi}\left(\eta_c^{\phi}-\eta_c^{\chi}\right)-6\eta_c^{\phi}\eta_c^{\chi}+4\left.\eta_c^{\chi}\right.^2-\left.\xi_c^{\chi}\right.^2\right)-\left.\epsilon_c^{\phi}\right.^3\left(48\left.\epsilon_c^{\chi}\right.^2-26\epsilon_c^{\chi}\eta_c^{\chi}+2\left.\eta_c^{\chi}\right.^2+\left.\xi_c^{\phi}\right.^2\right)\Bigg)\Bigg]\, \label{thirdderivativese}.
\end{multline}
$\eta^{ss}$ is given in (\ref{etassexp}):
\begin{equation}
\eta^{ss}\equiv\frac{\epsilon^{\chi}\eta^{\phi}-4\epsilon^{\phi}\epsilon^{\chi}+\epsilon^{\phi}\eta^{\chi}}{\epsilon} \, .\tag{\ref{etassexp}}
\end{equation}
As above, the $\delta N$ equations then give the observables of interest:
\begin{equation}
\frac{6}{5}f_{\mathrm{NL}}^{(4)}=\frac{1}{2}\frac{\frac{4\epsilon_c^{\phi}\epsilon_c^{\chi}}{\epsilon_c^2}\eta_c^{ss}\left(\frac{x_e}{\epsilon_*^{\phi}}-\frac{x_e}{\epsilon_*^{\chi}}\right)^2-x_e^3\left(\frac{\eta_*^{\phi}-2\epsilon_*^{\phi}}{\left(\epsilon_*^{\phi}\right)^2}\right)-y_e^3\left(\frac{\eta_*^{\chi}-2\epsilon_*^{\chi}}{\left(\epsilon_*^{\chi}\right)^2}\right)}{\left(\frac{x_e^2}{2\epsilon_*^{\phi}}+\frac{y_e^2}{2\epsilon_*^{\chi}}\right)^2}\, ,
\end{equation}
\begin{multline}
\tau_{\mathrm{NL}}=\frac{4}{\left(y_e^2\epsilon_*^{\phi}+x_e^2\epsilon_*^{\chi}\right)^3}\Bigg[\left(x_e^4\left.\epsilon_*^{\chi}\right.^3\left(\eta_*^{\phi}-2\epsilon_*^{\phi}\right)^2+y_e^4\left.\epsilon_*^{\phi}\right.^3\left(\eta_*^{\chi}-2\epsilon_*^{\chi}\right)^2\right)\\
+8\frac{\epsilon_c^{\chi}\epsilon_c^{\phi}}{\epsilon_c^2}\eta_c^{ss}\left(y_e\epsilon_*^{\phi}-x_e\epsilon_*^{\chi}\right)\left(x_e^2\left.\epsilon_*^{\chi}\right.^2\left(\eta_*^{\phi}-2\epsilon_*^{\phi}\right)+y_e^2\left.\epsilon_*^{\phi}\right.^2\left(\eta_*^{\chi}-2\epsilon_*^{\chi}\right)\right)\\
+16\left(\frac{\epsilon_c^{\chi}\epsilon_c^{\phi}}{\epsilon_c^2}\right)^2\left.\eta_c^{ss}\right.^2\left(y_e\epsilon_*^{\phi}-x_e\epsilon_*^{\chi}\right)^2\left(\epsilon_*^{\phi}+\epsilon_*^{\chi}\right)\Bigg]\, ,
\end{multline}
and
\begin{multline}
g_{\mathrm{NL}}=\frac{100}{27}\frac{1}{\left(y_e^2\epsilon_*^{\phi}+x_e^2\epsilon_*^{\chi}\right)^3}\Bigg[-\frac{1}{\epsilon_c}\left(x_e^3\epsilon_c^{\phi}\left.\epsilon_*^{\chi}\right.^3\left(2\left(\epsilon_*^{\phi}-\eta_*^{\phi}\right)\eta_*^{\phi}+\left.\xi_*^{\phi}\right.^2\right)+y_e^3\epsilon_c^{\chi}\left.\epsilon_*^{\phi}\right.^3\left(2\left(\epsilon_*^{\chi}-\eta_*^{\chi}\right)\eta_*^{\chi}+\left.\xi_*^{\chi}\right.^2\right)\right)\\
+6\frac{\epsilon_c^{\chi}\epsilon_c^{\phi}}{\epsilon_c^2}\eta_c^{ss}\left(y_e\epsilon_*^{\phi}-x_e\epsilon_*^{\chi}\right)\left(x_e^2\left.\epsilon_*^{\chi}\right.^2\left(\eta_*^{\phi}-2\epsilon_*^{\phi}\right)+y_e^2\left.\epsilon_*^{\phi}\right.^2\left(\eta_*^{\chi}-2\epsilon_*^{\chi}\right)\right)\\
+2\frac{\epsilon_c^{\chi}\epsilon_c^{\phi}}{\epsilon_c^5}\left(y_e\epsilon_*^{\phi}-x_e\epsilon_*^{\chi}\right)^3\Bigg(\epsilon_c^{\phi}\left.\epsilon_c^{\chi}\right.^2\left(26\epsilon_c^{\chi}\eta_c^{\phi}+4\left.\eta_c^{\phi}\right.^2-6\eta_c^{\phi}\eta_c^{\chi}-\left.\xi_c^{\phi}\right.^2\right)-\left.\epsilon_c^{\chi}\right.^3\left(2\left.\eta_c^{\phi}\right.^2+\left.\xi_c^{\phi}\right.^2\right)\\
+\left.\epsilon_c^{\phi}\right.^3\left(48\left.\epsilon_c^{\chi}\right.^2-26\epsilon_c^{\chi}\eta_c^{\chi}+2\left.\eta_c^{\chi}\right.^2+\left.\xi_c^{\chi}\right.^2\right)+\left.\epsilon_c^{\phi}\right.^2\epsilon_c^{\chi}\left(-48\left.\epsilon_c^{\chi}\right.^2+6\eta_c^{\phi}\eta_c^{\chi}-4\left.\eta_c^{\chi}\right.^2+22\epsilon_c^{\chi}\left(-\eta_c^{\phi}+\eta_c^{\chi}\right)+\left.\xi_c^{\chi}\right.^2\right)\Bigg)\Bigg]\, .
\end{multline}
Suppressing the slow-roll terms gives the expressions in (\ref{fnlexp}), (\ref{taunlexp}) and (\ref{gnlexp}).
\end{widetext}

\section{Five- and Six-Point Non-Linearity Parameters}\label{fiveandsix}
In this appendix we give explicit expressions for the non-linearity parameters which describe the amplitude of the various local limits of the five- and six-point functions assuming independent gaussian scalar field fluctuations at horizon exit.  By using the general expression (\ref{npointparameters}) we obtain for the five-point non-linearity parameters:
\begin{align}
F_{\mathrm{NL},1}^{(5)}&=\frac{\sum_{IJKL}N_{,IJ}N_{,JK}N_{,KL}N_{,I}N_{,L}}{\left(\sum_MN_{,M}^2\right)^4}\, ,\nonumber \\
F_{\mathrm{NL},2}^{(5)}&=\frac{\sum_{IJKL}N_{,IJK}N_{,KL}N_{,I}N_{,J}N_{,L}}{\left(\sum_MN_{,M}^2\right)^4} \, ,\nonumber \\
F_{\mathrm{NL},3}^{(5)}&=\frac{\sum_{IJKL}N_{,IJKL}N_{,I}N_{,J}N_{,K}N_{,L}}{\left(\sum_MN_{,M}^2\right)^4}\, .
\end{align}
And for the six-point non-linearity parameters:
\begin{align}
F_{\mathrm{NL},1}^{(6)}&=\frac{\sum_{IJKLM}N_{,IJ}N_{,JK}N_{,KL}N_{,LM}N_{,I}N_{,M}}{\left(\sum_NN_{,N}^2\right)^5}\, ,\nonumber \\
F_{\mathrm{NL},2}^{(6)}&=\frac{\sum_{IJKLM}N_{,IJK}N_{,KL}N_{,LM}N_{,I}N_{,J}N_{,M}}{\left(\sum_NN_{,N}^2\right)^5}\, ,\nonumber \\
F_{\mathrm{NL},3}^{(6)}&=\frac{\sum_{IJKLM}N_{,IJK}N_{,JL}N_{,KM}N_{,I}N_{,L}N_{,M}}{\left(\sum_NN_{,N}^2\right)^5}\, ,\nonumber \\
F_{\mathrm{NL},4}^{(6)}&=\frac{\sum_{IJKLM}N_{,IJK}N_{,KLM}N_{,I}N_{,J}N_{,L}N_{,M}}{\left(\sum_NN_{,N}^2\right)^5}\, ,\nonumber \\
F_{\mathrm{NL},5}^{(6)}&=\frac{\sum_{IJKLM}N_{,IJKL}N_{,LM}N_{,I}N_{,J}N_{,K}N_{,M}}{\left(\sum_NN_{,N}^2\right)^5}\, ,\nonumber \\
F_{\mathrm{NL},6}^{(6)}&=\frac{\sum_{IJKLM}N_{,IJKLM}N_{,I}N_{,J}N_{,K}N_{,L}N_{,M}}{\left(\sum_NN_{,N}^2\right)^5}\, .
\end{align}

It can easily be shown that there are many other consistency relations like the one found by Suyama and Yamaguchi \cite{Suyama:2007bg,Sugiyama:2011jt} which relate the non-linearity parameters for higher point functions.  For example, if we take
\begin{align}
	c_I&=\frac{\sum_{JK}N_{,IJK}N_{,J}N_{,K}}{\left[\sum_LN_{,L}^2\right]^{5/2}}\, , \nonumber \\
	d_I&=\frac{N_I}{\left[\sum_LN_{,L}^2\right]^{1/2}}\, ,
\end{align}
then the Cauchy-Schwarz inequality
\begin{equation}
 \left(\sum_Ic_I^2\right)\left(\sum_Jd_J^2\right)\geq\left(\sum_Ic_Id_I\right)^2
\end{equation}
implies that
\begin{equation}
	F_{\mathrm{NL},4}^{(6)}\geq\left(\frac{54g_{\mathrm{NL}}}{25}\right)^2\, .
\end{equation}
	
\bibliography{nongauss2}

\begin{thebibliography}{10}%
\makeatletter
\providecommand \@ifxundefined [1]{%
 \ifx #1\undefined \expandafter \@firstoftwo
 \else \expandafter \@secondoftwo
\fi
}%
\providecommand \@ifnum [1]{%
 \ifnum #1\expandafter \@firstoftwo
 \else \expandafter \@secondoftwo
\fi
}%
\providecommand \enquote [1]{``#1''}%
\providecommand \bibnamefont  [1]{#1}%
\providecommand \bibfnamefont [1]{#1}%
\providecommand \citenamefont [1]{#1}%
\providecommand\href[0]{\@sanitize\@href}%
\providecommand\@href[1]{\endgroup\@@startlink{#1}\endgroup\@@href}%
\providecommand\@@href[1]{#1\@@endlink}%
\providecommand \@sanitize [0]{\begingroup\catcode`\&12\catcode`\#12\relax}%
\@ifxundefined \pdfoutput {\@firstoftwo}{%
 \@ifnum{\z@=\pdfoutput}{\@firstoftwo}{\@secondoftwo}%
}{%
 \providecommand\@@startlink[1]{\leavevmode\special{html:<a href="#1">}}%
 \providecommand\@@endlink[0]{\special{html:</a>}}%
}{%
 \providecommand\@@startlink[1]{%
  \leavevmode
  \pdfstartlink
   attr{/Border[0 0 1 ]/H/I/C[0 1 1]}%
   user{/Subtype/Link/A<</Type/Action/S/URI/URI(#1)>>}%
  \relax
 }%
 \providecommand\@@endlink[0]{\pdfendlink}%
}%
\providecommand \url  [0]{\begingroup\@sanitize \@url }%
\providecommand \@url [1]{\endgroup\@href {#1}{\urlprefix}}%
\providecommand \urlprefix [0]{URL }%
\providecommand \Eprint[0]{\href }%
\@ifxundefined \urlstyle {%
  \providecommand \doi [1]{doi:\discretionary{}{}{}#1}%
}{%
  \providecommand \doi [0]{doi:\discretionary{}{}{}\begingroup
  \urlstyle{rm}\Url }%
}%
\providecommand \doibase [0]{http://dx.doi.org/}%
\providecommand \Doi[1]{\href{\doibase#1}}%
\providecommand \bibAnnote [3]{%
  \BibitemShut{#1}%
  \begin{quotation}\noindent
    \textsc{Key:}\ #2\\\textsc{Annotation:}\ #3%
  \end{quotation}%
}%
\providecommand \bibAnnoteFile [2]{%
  \IfFileExists{#2}{\bibAnnote {#1} {#2} {\input{#2}}}{}%
}%
\providecommand \typeout [0]{\immediate \write \m@ne }%
\providecommand \selectlanguage [0]{\@gobble}%
\providecommand \bibinfo [0]{\@secondoftwo}%
\providecommand \bibfield [0]{\@secondoftwo}%
\providecommand \translation [1]{[#1]}%
\providecommand \BibitemOpen[0]{}%
\providecommand \bibitemStop [0]{}%
\providecommand \bibitemNoStop [0]{.\EOS\space}%
\providecommand \EOS [0]{\spacefactor3000\relax}%
\providecommand \BibitemShut [1]{\csname bibitem#1\endcsname}%
\bibitem{Meyers:2010rg}%
  \BibitemOpen
  \bibfield{author}{%
  \bibinfo {author} {\bibfnamefont{J.}~\bibnamefont{Meyers}}\ and\ \bibinfo
  {author} {\bibfnamefont{N.}~\bibnamefont{Sivanandam}}}%
   (\bibinfo {year} {2010}),\ \bibinfo {note} {* Temporary entry *},\
  \Eprint{http://arxiv.org/abs/1011.4934}{arXiv:1011.4934 [astro-ph.CO]}%
  \bibAnnoteFile{NoStop}{Meyers:2010rg}%
\bibitem{Komatsu:2010fb}%
  \BibitemOpen
  \bibfield{author}{%
  \bibinfo {author} {\bibfnamefont{E.}~\bibnamefont{Komatsu}}, \bibinfo
  {author} {\bibfnamefont{K.}~\bibnamefont{Smith}}, \bibinfo {author}
  {\bibfnamefont{J.}~\bibnamefont{Dunkley}}, \bibinfo {author}
  {\bibfnamefont{C.}~\bibnamefont{Bennett}}, \bibinfo {author}
  {\bibfnamefont{B.}~\bibnamefont{Gold}}, \emph{et~al.}}%
   (\bibinfo {year} {2010}),\
  \Eprint{http://arxiv.org/abs/arXiv:1001.4538}{arXiv:arXiv:1001.4538
  [astro-ph.CO]}%
  \bibAnnoteFile{NoStop}{Komatsu:2010fb}%
\bibitem{Mukhanov:1981xt}%
  \BibitemOpen
  \bibfield{author}{%
  \bibinfo {author} {\bibfnamefont{V.~F.}\ \bibnamefont{Mukhanov}}\ and\
  \bibinfo {author} {\bibfnamefont{G.~V.}\ \bibnamefont{Chibisov}},\ }%
  \bibfield{journal}{%
  \bibinfo {journal} {JETP Lett.}\ }%
  \textbf{\bibinfo {volume} {33}},\ \bibinfo {pages} {532} (\bibinfo {year}
  {1981})%
  \bibAnnoteFile{NoStop}{Mukhanov:1981xt}%
\bibitem{Hawking:1982cz}%
  \BibitemOpen
  \bibfield{author}{%
  \bibinfo {author} {\bibfnamefont{S.~W.}\ \bibnamefont{Hawking}},\ }%
  \bibfield{journal}{%
  \Doi{10.1016/0370-2693(82)90373-2}{\bibinfo {journal} {Phys. Lett.}}\ }%
  \textbf{\bibinfo {volume} {B115}},\ \bibinfo {pages} {295} (\bibinfo {year}
  {1982})%
  \bibAnnoteFile{NoStop}{Hawking:1982cz}%
\bibitem{Starobinsky:1982ee}%
  \BibitemOpen
  \bibfield{author}{%
  \bibinfo {author} {\bibfnamefont{A.~A.}\ \bibnamefont{Starobinsky}},\ }%
  \bibfield{journal}{%
  \Doi{10.1016/0370-2693(82)90541-X}{\bibinfo {journal} {Phys. Lett.}}\ }%
  \textbf{\bibinfo {volume} {B117}},\ \bibinfo {pages} {175} (\bibinfo {year}
  {1982})%
  \bibAnnoteFile{NoStop}{Starobinsky:1982ee}%
\bibitem{Guth:1982ec}%
  \BibitemOpen
  \bibfield{author}{%
  \bibinfo {author} {\bibfnamefont{A.~H.}\ \bibnamefont{Guth}}\ and\ \bibinfo
  {author} {\bibfnamefont{S.~Y.}\ \bibnamefont{Pi}},\ }%
  \bibfield{journal}{%
  \Doi{10.1103/PhysRevLett.49.1110}{\bibinfo {journal} {Phys. Rev. Lett.}}\ }%
  \textbf{\bibinfo {volume} {49}},\ \bibinfo {pages} {1110} (\bibinfo {year}
  {1982})%
  \bibAnnoteFile{NoStop}{Guth:1982ec}%
\bibitem{Bardeen:1983qw}%
  \BibitemOpen
  \bibfield{author}{%
  \bibinfo {author} {\bibfnamefont{J.~M.}\ \bibnamefont{Bardeen}}, \bibinfo
  {author} {\bibfnamefont{P.~J.}\ \bibnamefont{Steinhardt}},\ and\ \bibinfo
  {author} {\bibfnamefont{M.~S.}\ \bibnamefont{Turner}},\ }%
  \bibfield{journal}{%
  \Doi{10.1103/PhysRevD.28.679}{\bibinfo {journal} {Phys. Rev.}}\ }%
  \textbf{\bibinfo {volume} {D28}},\ \bibinfo {pages} {679} (\bibinfo {year}
  {1983})%
  \bibAnnoteFile{NoStop}{Bardeen:1983qw}%
\bibitem{Fischler:1985ky}%
  \BibitemOpen
  \bibfield{author}{%
  \bibinfo {author} {\bibfnamefont{W.}~\bibnamefont{Fischler}}, \bibinfo
  {author} {\bibfnamefont{B.}~\bibnamefont{Ratra}},\ and\ \bibinfo {author}
  {\bibfnamefont{L.}~\bibnamefont{Susskind}},\ }%
  \bibfield{journal}{%
  \Doi{10.1016/0550-3213(85)90011-2}{\bibinfo {journal} {Nucl. Phys.}}\ }%
  \textbf{\bibinfo {volume} {B259}},\ \bibinfo {pages} {730} (\bibinfo {year}
  {1985})%
  \bibAnnoteFile{NoStop}{Fischler:1985ky}%
\bibitem{Seery:2005gb}%
  \BibitemOpen
  \bibfield{author}{%
  \bibinfo {author} {\bibfnamefont{D.}~\bibnamefont{Seery}}\ and\ \bibinfo
  {author} {\bibfnamefont{J.~E.}\ \bibnamefont{Lidsey}},\ }%
  \bibfield{journal}{%
  \Doi{10.1088/1475-7516/2005/09/011}{\bibinfo {journal} {JCAP}}\ }%
  \textbf{\bibinfo {volume} {0509}},\ \bibinfo {pages} {011} (\bibinfo {year}
  {2005}),\
  \Eprint{http://arxiv.org/abs/astro-ph/0506056}{arXiv:astro-ph/0506056
  [astro-ph]}%
  \bibAnnoteFile{NoStop}{Seery:2005gb}%
\bibitem{Gao:2008dt}%
  \BibitemOpen
  \bibfield{author}{%
  \bibinfo {author} {\bibfnamefont{X.}~\bibnamefont{Gao}},\ }%
  \bibfield{journal}{%
  \Doi{10.1088/1475-7516/2008/06/029}{\bibinfo {journal} {JCAP}}\ }%
  \textbf{\bibinfo {volume} {0806}},\ \bibinfo {pages} {029} (\bibinfo {year}
  {2008}),\ \bibinfo {note} {* Brief entry *},\
  \Eprint{http://arxiv.org/abs/arXiv:0804.1055}{arXiv:arXiv:0804.1055
  [astro-ph]}%
  \bibAnnoteFile{NoStop}{Gao:2008dt}%
\bibitem{Maldacena:2002vr}%
  \BibitemOpen
  \bibfield{author}{%
  \bibinfo {author} {\bibfnamefont{J.~M.}\ \bibnamefont{Maldacena}},\ }%
  \bibfield{journal}{%
  \bibinfo {journal} {JHEP}\ }%
  \textbf{\bibinfo {volume} {0305}},\ \bibinfo {pages} {013} (\bibinfo {year}
  {2003}),\
  \Eprint{http://arxiv.org/abs/astro-ph/0210603}{arXiv:astro-ph/0210603
  [astro-ph]}%
  \bibAnnoteFile{NoStop}{Maldacena:2002vr}%
\bibitem{Seery:2005wm}%
  \BibitemOpen
  \bibfield{author}{%
  \bibinfo {author} {\bibfnamefont{D.}~\bibnamefont{Seery}}\ and\ \bibinfo
  {author} {\bibfnamefont{J.~E.}\ \bibnamefont{Lidsey}},\ }%
  \bibfield{journal}{%
  \Doi{10.1088/1475-7516/2005/06/003}{\bibinfo {journal} {JCAP}}\ }%
  \textbf{\bibinfo {volume} {0506}},\ \bibinfo {pages} {003} (\bibinfo {year}
  {2005}),\
  \Eprint{http://arxiv.org/abs/astro-ph/0503692}{arXiv:astro-ph/0503692
  [astro-ph]}%
  \bibAnnoteFile{NoStop}{Seery:2005wm}%
\bibitem{Linde:1996gt}%
  \BibitemOpen
  \bibfield{author}{%
  \bibinfo {author} {\bibfnamefont{A.~D.}\ \bibnamefont{Linde}}\ and\ \bibinfo
  {author} {\bibfnamefont{V.~F.}\ \bibnamefont{Mukhanov}},\ }%
  \bibfield{journal}{%
  \Doi{10.1103/PhysRevD.56.R535}{\bibinfo {journal} {Phys. Rev.}}\ }%
  \textbf{\bibinfo {volume} {D56}},\ \bibinfo {pages} {535} (\bibinfo {year}
  {1997}),\
  \Eprint{http://arxiv.org/abs/astro-ph/9610219}{arXiv:astro-ph/9610219}%
  \bibAnnoteFile{NoStop}{Linde:1996gt}%
\bibitem{Bernardeau:2002jy}%
  \BibitemOpen
  \bibfield{author}{%
  \bibinfo {author} {\bibfnamefont{F.}~\bibnamefont{Bernardeau}}\ and\ \bibinfo
  {author} {\bibfnamefont{J.-P.}\ \bibnamefont{Uzan}},\ }%
  \bibfield{journal}{%
  \Doi{10.1103/PhysRevD.66.103506}{\bibinfo {journal} {Phys. Rev.}}\ }%
  \textbf{\bibinfo {volume} {D66}},\ \bibinfo {pages} {103506} (\bibinfo {year}
  {2002}),\ \Eprint{http://arxiv.org/abs/hep-ph/0207295}{arXiv:hep-ph/0207295}%
  \bibAnnoteFile{NoStop}{Bernardeau:2002jy}%
\bibitem{Bernardeau:2002jf}%
  \BibitemOpen
  \bibfield{author}{%
  \bibinfo {author} {\bibfnamefont{F.}~\bibnamefont{Bernardeau}}\ and\ \bibinfo
  {author} {\bibfnamefont{J.-P.}\ \bibnamefont{Uzan}},\ }%
  \bibfield{journal}{%
  \Doi{10.1103/PhysRevD.67.121301}{\bibinfo {journal} {Phys. Rev.}}\ }%
  \textbf{\bibinfo {volume} {D67}},\ \bibinfo {pages} {121301} (\bibinfo {year}
  {2003}),\
  \Eprint{http://arxiv.org/abs/astro-ph/0209330}{arXiv:astro-ph/0209330}%
  \bibAnnoteFile{NoStop}{Bernardeau:2002jf}%
\bibitem{Lyth:2002my}%
  \BibitemOpen
  \bibfield{author}{%
  \bibinfo {author} {\bibfnamefont{D.~H.}\ \bibnamefont{Lyth}}, \bibinfo
  {author} {\bibfnamefont{C.}~\bibnamefont{Ungarelli}},\ and\ \bibinfo {author}
  {\bibfnamefont{D.}~\bibnamefont{Wands}},\ }%
  \bibfield{journal}{%
  \Doi{10.1103/PhysRevD.67.023503}{\bibinfo {journal} {Phys. Rev.}}\ }%
  \textbf{\bibinfo {volume} {D67}},\ \bibinfo {pages} {023503} (\bibinfo {year}
  {2003}),\
  \Eprint{http://arxiv.org/abs/astro-ph/0208055}{arXiv:astro-ph/0208055}%
  \bibAnnoteFile{NoStop}{Lyth:2002my}%
\bibitem{Byrnes:2009qy}%
  \BibitemOpen
  \bibfield{author}{%
  \bibinfo {author} {\bibfnamefont{C.~T.}\ \bibnamefont{Byrnes}}\ and\ \bibinfo
  {author} {\bibfnamefont{G.}~\bibnamefont{Tasinato}},\ }%
  \bibfield{journal}{%
  \Doi{10.1088/1475-7516/2009/08/016}{\bibinfo {journal} {JCAP}}\ }%
  \textbf{\bibinfo {volume} {0908}},\ \bibinfo {pages} {016} (\bibinfo {year}
  {2009}),\ \bibinfo {note} {* Brief entry *},\
  \Eprint{http://arxiv.org/abs/arXiv:0906.0767}{arXiv:arXiv:0906.0767
  [astro-ph.CO]}%
  \bibAnnoteFile{NoStop}{Byrnes:2009qy}%
\bibitem{Battefeld:2006sz}%
  \BibitemOpen
  \bibfield{author}{%
  \bibinfo {author} {\bibfnamefont{T.}~\bibnamefont{Battefeld}}\ and\ \bibinfo
  {author} {\bibfnamefont{R.}~\bibnamefont{Easther}},\ }%
  \bibfield{journal}{%
  \Doi{10.1088/1475-7516/2007/03/020}{\bibinfo {journal} {JCAP}}\ }%
  \textbf{\bibinfo {volume} {0703}},\ \bibinfo {pages} {020} (\bibinfo {year}
  {2007}),\
  \Eprint{http://arxiv.org/abs/astro-ph/0610296}{arXiv:astro-ph/0610296}%
  \bibAnnoteFile{NoStop}{Battefeld:2006sz}%
\bibitem{Chen:2009we}%
  \BibitemOpen
  \bibfield{author}{%
  \bibinfo {author} {\bibfnamefont{X.}~\bibnamefont{Chen}}\ and\ \bibinfo
  {author} {\bibfnamefont{Y.}~\bibnamefont{Wang}},\ }%
  \bibfield{journal}{%
  \Doi{10.1103/PhysRevD.81.063511}{\bibinfo {journal} {Phys.Rev.}}\ }%
  \textbf{\bibinfo {volume} {D81}},\ \bibinfo {pages} {063511} (\bibinfo {year}
  {2010}),\ \Eprint{http://arxiv.org/abs/arXiv:0909.0496}{arXiv:arXiv:0909.0496
  [astro-ph.CO]}%
  \bibAnnoteFile{NoStop}{Chen:2009we}%
\bibitem{Sasaki:2008uc}%
  \BibitemOpen
  \bibfield{author}{%
  \bibinfo {author} {\bibfnamefont{M.}~\bibnamefont{Sasaki}},\ }%
  \bibfield{journal}{%
  \Doi{10.1143/PTP.120.159}{\bibinfo {journal} {Prog.Theor.Phys.}}\ }%
  \textbf{\bibinfo {volume} {120}},\ \bibinfo {pages} {159} (\bibinfo {year}
  {2008}),\ \bibinfo {note} {* Brief entry *},\
  \Eprint{http://arxiv.org/abs/arXiv:0805.0974}{arXiv:arXiv:0805.0974
  [astro-ph]}%
  \bibAnnoteFile{NoStop}{Sasaki:2008uc}%
\bibitem{Battefeld:2009ym}%
  \BibitemOpen
  \bibfield{author}{%
  \bibinfo {author} {\bibfnamefont{D.}~\bibnamefont{Battefeld}}\ and\ \bibinfo
  {author} {\bibfnamefont{T.}~\bibnamefont{Battefeld}},\ }%
  \bibfield{journal}{%
  \Doi{10.1088/1475-7516/2009/11/010}{\bibinfo {journal} {JCAP}}\ }%
  \textbf{\bibinfo {volume} {0911}},\ \bibinfo {pages} {010} (\bibinfo {year}
  {2009}),\ \Eprint{http://arxiv.org/abs/arXiv:0908.4269}{arXiv:arXiv:0908.4269
  [hep-th]}%
  \bibAnnoteFile{NoStop}{Battefeld:2009ym}%
\bibitem{Byrnes:2008zy}%
  \BibitemOpen
  \bibfield{author}{%
  \bibinfo {author} {\bibfnamefont{C.~T.}\ \bibnamefont{Byrnes}}, \bibinfo
  {author} {\bibfnamefont{K.-Y.}\ \bibnamefont{Choi}},\ and\ \bibinfo {author}
  {\bibfnamefont{L.~M.}\ \bibnamefont{Hall}},\ }%
  \bibfield{journal}{%
  \Doi{10.1088/1475-7516/2009/02/017}{\bibinfo {journal} {JCAP}}\ }%
  \textbf{\bibinfo {volume} {0902}},\ \bibinfo {pages} {017} (\bibinfo {year}
  {2009}),\ \bibinfo {note} {* Brief entry *},\
  \Eprint{http://arxiv.org/abs/arXiv:0812.0807}{arXiv:arXiv:0812.0807
  [astro-ph]}%
  \bibAnnoteFile{NoStop}{Byrnes:2008zy}%
\bibitem{Cai:2009hw}%
  \BibitemOpen
  \bibfield{author}{%
  \bibinfo {author} {\bibfnamefont{Y.-F.}\ \bibnamefont{Cai}}\ and\ \bibinfo
  {author} {\bibfnamefont{H.-Y.}\ \bibnamefont{Xia}},\ }%
  \bibfield{journal}{%
  \Doi{10.1016/j.physletb.2009.05.047}{\bibinfo {journal} {Phys.Lett.}}\ }%
  \textbf{\bibinfo {volume} {B677}},\ \bibinfo {pages} {226} (\bibinfo {year}
  {2009}),\ \Eprint{http://arxiv.org/abs/0904.0062}{arXiv:0904.0062 [hep-th]}%
  \bibAnnoteFile{NoStop}{Cai:2009hw}%
\bibitem{Dvali:2003em}%
  \BibitemOpen
  \bibfield{author}{%
  \bibinfo {author} {\bibfnamefont{G.}~\bibnamefont{Dvali}}, \bibinfo {author}
  {\bibfnamefont{A.}~\bibnamefont{Gruzinov}},\ and\ \bibinfo {author}
  {\bibfnamefont{M.}~\bibnamefont{Zaldarriaga}},\ }%
  \bibfield{journal}{%
  \Doi{10.1103/PhysRevD.69.023505}{\bibinfo {journal} {Phys. Rev.}}\ }%
  \textbf{\bibinfo {volume} {D69}},\ \bibinfo {pages} {023505} (\bibinfo {year}
  {2004}),\
  \Eprint{http://arxiv.org/abs/astro-ph/0303591}{arXiv:astro-ph/0303591}%
  \bibAnnoteFile{NoStop}{Dvali:2003em}%
\bibitem{Kofman:2003nx}%
  \BibitemOpen
  \bibfield{author}{%
  \bibinfo {author} {\bibfnamefont{L.}~\bibnamefont{Kofman}}}%
   (\bibinfo {year} {2003}),\
  \Eprint{http://arxiv.org/abs/astro-ph/0303614}{arXiv:astro-ph/0303614}%
  \bibAnnoteFile{NoStop}{Kofman:2003nx}%
\bibitem{Rigopoulos:2005ae}%
  \BibitemOpen
  \bibfield{author}{%
  \bibinfo {author} {\bibfnamefont{G.}~\bibnamefont{Rigopoulos}}, \bibinfo
  {author} {\bibfnamefont{E.}~\bibnamefont{Shellard}},\ and\ \bibinfo {author}
  {\bibfnamefont{B.}~\bibnamefont{van Tent}},\ }%
  \bibfield{journal}{%
  \Doi{10.1103/PhysRevD.73.083522}{\bibinfo {journal} {Phys.Rev.}}\ }%
  \textbf{\bibinfo {volume} {D73}},\ \bibinfo {pages} {083522} (\bibinfo {year}
  {2006}),\
  \Eprint{http://arxiv.org/abs/astro-ph/0506704}{arXiv:astro-ph/0506704
  [astro-ph]}%
  \bibAnnoteFile{NoStop}{Rigopoulos:2005ae}%
\bibitem{Rigopoulos:2005us}%
  \BibitemOpen
  \bibfield{author}{%
  \bibinfo {author} {\bibfnamefont{G.}~\bibnamefont{Rigopoulos}}, \bibinfo
  {author} {\bibfnamefont{E.}~\bibnamefont{Shellard}},\ and\ \bibinfo {author}
  {\bibfnamefont{B.}~\bibnamefont{van Tent}},\ }%
  \bibfield{journal}{%
  \Doi{10.1103/PhysRevD.76.083512}{\bibinfo {journal} {Phys.Rev.}}\ }%
  \textbf{\bibinfo {volume} {D76}},\ \bibinfo {pages} {083512} (\bibinfo {year}
  {2007}),\
  \Eprint{http://arxiv.org/abs/astro-ph/0511041}{arXiv:astro-ph/0511041
  [astro-ph]}%
  \bibAnnoteFile{NoStop}{Rigopoulos:2005us}%
\bibitem{Byrnes:2010em}%
  \BibitemOpen
  \bibfield{author}{%
  \bibinfo {author} {\bibfnamefont{C.~T.}\ \bibnamefont{Byrnes}}\ and\ \bibinfo
  {author} {\bibfnamefont{K.-Y.}\ \bibnamefont{Choi}},\ }%
  \bibfield{journal}{%
  \bibinfo {journal} {Adv. Astron.}\ }%
  \textbf{\bibinfo {volume} {2010}},\ \bibinfo {pages} {724525} (\bibinfo
  {year} {2010}),\ \Eprint{http://arxiv.org/abs/1002.3110}{arXiv:1002.3110
  [astro-ph.CO]}%
  \bibAnnoteFile{NoStop}{Byrnes:2010em}%
\bibitem{Wands:2010af}%
  \BibitemOpen
  \bibfield{author}{%
  \bibinfo {author} {\bibfnamefont{D.}~\bibnamefont{Wands}},\ }%
  \bibfield{journal}{%
  \Doi{10.1088/0264-9381/27/12/124002}{\bibinfo {journal} {Class.Quant.Grav.}}\
  }%
  \textbf{\bibinfo {volume} {27}},\ \bibinfo {pages} {124002} (\bibinfo {year}
  {2010}),\ \Eprint{http://arxiv.org/abs/1004.0818}{arXiv:1004.0818
  [astro-ph.CO]}%
  \bibAnnoteFile{NoStop}{Wands:2010af}%
\bibitem{Weinberg:2008zzc}%
  \BibitemOpen
  \bibfield{author}{%
  \bibinfo {author} {\bibfnamefont{S.}~\bibnamefont{Weinberg}},\ }%
  \emph{\bibinfo {title} {Cosmology}}\ (\bibinfo {publisher} {Oxford University
  Press},\ \bibinfo {year} {2008})%
  \bibAnnoteFile{NoStop}{Weinberg:2008zzc}%
\bibitem{Weinberg:2003sw}%
  \BibitemOpen
  \bibfield{author}{%
  \bibinfo {author} {\bibfnamefont{S.}~\bibnamefont{Weinberg}},\ }%
  \bibfield{journal}{%
  \Doi{10.1103/PhysRevD.67.123504}{\bibinfo {journal} {Phys.Rev.}}\ }%
  \textbf{\bibinfo {volume} {D67}},\ \bibinfo {pages} {123504} (\bibinfo {year}
  {2003}),\
  \Eprint{http://arxiv.org/abs/astro-ph/0302326}{arXiv:astro-ph/0302326
  [astro-ph]}%
  \bibAnnoteFile{NoStop}{Weinberg:2003sw}%
\bibitem{Weinberg:2004kr}%
  \BibitemOpen
  \bibfield{author}{%
  \bibinfo {author} {\bibfnamefont{S.}~\bibnamefont{Weinberg}},\ }%
  \bibfield{journal}{%
  \Doi{10.1103/PhysRevD.70.043541}{\bibinfo {journal} {Phys.Rev.}}\ }%
  \textbf{\bibinfo {volume} {D70}},\ \bibinfo {pages} {043541} (\bibinfo {year}
  {2004}),\
  \Eprint{http://arxiv.org/abs/astro-ph/0401313}{arXiv:astro-ph/0401313
  [astro-ph]}%
  \bibAnnoteFile{NoStop}{Weinberg:2004kr}%
\bibitem{Weinberg:2008nf}%
  \BibitemOpen
  \bibfield{author}{%
  \bibinfo {author} {\bibfnamefont{S.}~\bibnamefont{Weinberg}},\ }%
  \bibfield{journal}{%
  \Doi{10.1103/PhysRevD.78.123521}{\bibinfo {journal} {Phys.Rev.}}\ }%
  \textbf{\bibinfo {volume} {D78}},\ \bibinfo {pages} {123521} (\bibinfo {year}
  {2008}),\ \Eprint{http://arxiv.org/abs/arXiv:0808.2909}{arXiv:arXiv:0808.2909
  [hep-th]}%
  \bibAnnoteFile{NoStop}{Weinberg:2008nf}%
\bibitem{Bucher:1999re}%
  \BibitemOpen
  \bibfield{author}{%
  \bibinfo {author} {\bibfnamefont{M.}~\bibnamefont{Bucher}}, \bibinfo {author}
  {\bibfnamefont{K.}~\bibnamefont{Moodley}},\ and\ \bibinfo {author}
  {\bibfnamefont{N.}~\bibnamefont{Turok}},\ }%
  \bibfield{journal}{%
  \Doi{10.1103/PhysRevD.62.083508}{\bibinfo {journal} {Phys. Rev.}}\ }%
  \textbf{\bibinfo {volume} {D62}},\ \bibinfo {pages} {083508} (\bibinfo {year}
  {2000}),\
  \Eprint{http://arxiv.org/abs/astro-ph/9904231}{arXiv:astro-ph/9904231}%
  \bibAnnoteFile{NoStop}{Bucher:1999re}%
\bibitem{Bucher:2000kb}%
  \BibitemOpen
  \bibfield{author}{%
  \bibinfo {author} {\bibfnamefont{M.}~\bibnamefont{Bucher}}, \bibinfo {author}
  {\bibfnamefont{K.}~\bibnamefont{Moodley}},\ and\ \bibinfo {author}
  {\bibfnamefont{N.}~\bibnamefont{Turok}},\ }%
  \bibfield{journal}{%
  \Doi{10.1103/PhysRevD.66.023528}{\bibinfo {journal} {Phys. Rev.}}\ }%
  \textbf{\bibinfo {volume} {D66}},\ \bibinfo {pages} {023528} (\bibinfo {year}
  {2002}),\
  \Eprint{http://arxiv.org/abs/astro-ph/0007360}{arXiv:astro-ph/0007360}%
  \bibAnnoteFile{NoStop}{Bucher:2000kb}%
\bibitem{Bucher:2000cd}%
  \BibitemOpen
  \bibfield{author}{%
  \bibinfo {author} {\bibfnamefont{M.}~\bibnamefont{Bucher}}, \bibinfo {author}
  {\bibfnamefont{K.}~\bibnamefont{Moodley}},\ and\ \bibinfo {author}
  {\bibfnamefont{N.}~\bibnamefont{Turok}}}%
   (\bibinfo {year} {2000}),\
  \Eprint{http://arxiv.org/abs/astro-ph/0011025}{arXiv:astro-ph/0011025}%
  \bibAnnoteFile{NoStop}{Bucher:2000cd}%
\bibitem{Bucher:2000hy}%
  \BibitemOpen
  \bibfield{author}{%
  \bibinfo {author} {\bibfnamefont{M.}~\bibnamefont{Bucher}}, \bibinfo {author}
  {\bibfnamefont{K.}~\bibnamefont{Moodley}},\ and\ \bibinfo {author}
  {\bibfnamefont{N.}~\bibnamefont{Turok}},\ }%
  \bibfield{journal}{%
  \Doi{10.1103/PhysRevLett.87.191301}{\bibinfo {journal} {Phys. Rev. Lett.}}\
  }%
  \textbf{\bibinfo {volume} {87}},\ \bibinfo {pages} {191301} (\bibinfo {year}
  {2001}),\
  \Eprint{http://arxiv.org/abs/astro-ph/0012141}{arXiv:astro-ph/0012141}%
  \bibAnnoteFile{NoStop}{Bucher:2000hy}%
\bibitem{Smidt:2010sv}%
  \BibitemOpen
  \bibfield{author}{%
  \bibinfo {author} {\bibfnamefont{J.}~\bibnamefont{Smidt}} \emph{et~al.}}%
   (\bibinfo {year} {2010}),\
  \Eprint{http://arxiv.org/abs/1001.5026}{arXiv:1001.5026 [astro-ph.CO]}%
  \bibAnnoteFile{NoStop}{Smidt:2010sv}%
\bibitem{Fergusson:2010gn}%
  \BibitemOpen
  \bibfield{author}{%
  \bibinfo {author} {\bibfnamefont{J.}~\bibnamefont{Fergusson}}, \bibinfo
  {author} {\bibfnamefont{D.}~\bibnamefont{Regan}},\ and\ \bibinfo {author}
  {\bibfnamefont{E.}~\bibnamefont{Shellard}}}%
   (\bibinfo {year} {2010}),\ \bibinfo {note} {* Temporary entry *},\
  \Eprint{http://arxiv.org/abs/1012.6039}{arXiv:1012.6039 [astro-ph.CO]}%
  \bibAnnoteFile{NoStop}{Fergusson:2010gn}%
\bibitem{Byrnes:2006vq}%
  \BibitemOpen
  \bibfield{author}{%
  \bibinfo {author} {\bibfnamefont{C.~T.}\ \bibnamefont{Byrnes}}, \bibinfo
  {author} {\bibfnamefont{M.}~\bibnamefont{Sasaki}},\ and\ \bibinfo {author}
  {\bibfnamefont{D.}~\bibnamefont{Wands}},\ }%
  \bibfield{journal}{%
  \Doi{10.1103/PhysRevD.74.123519}{\bibinfo {journal} {Phys. Rev.}}\ }%
  \textbf{\bibinfo {volume} {D74}},\ \bibinfo {pages} {123519} (\bibinfo {year}
  {2006}),\
  \Eprint{http://arxiv.org/abs/astro-ph/0611075}{arXiv:astro-ph/0611075}%
  \bibAnnoteFile{NoStop}{Byrnes:2006vq}%
\bibitem{Komatsu:2010hc}%
  \BibitemOpen
  \bibfield{author}{%
  \bibinfo {author} {\bibfnamefont{E.}~\bibnamefont{Komatsu}},\ }%
  \bibfield{journal}{%
  \Doi{10.1088/0264-9381/27/12/124010}{\bibinfo {journal} {Class.Quant.Grav.}}\
  }%
  \textbf{\bibinfo {volume} {27}},\ \bibinfo {pages} {124010} (\bibinfo {year}
  {2010}),\ \bibinfo {note} {* Temporary entry *},\
  \Eprint{http://arxiv.org/abs/arXiv:1003.6097}{arXiv:arXiv:1003.6097
  [astro-ph.CO]}%
  \bibAnnoteFile{NoStop}{Komatsu:2010hc}%
\bibitem{Jeong:2008rj}%
  \BibitemOpen
  \bibfield{author}{%
  \bibinfo {author} {\bibfnamefont{D.}~\bibnamefont{Jeong}}\ and\ \bibinfo
  {author} {\bibfnamefont{E.}~\bibnamefont{Komatsu}},\ }%
  \bibfield{journal}{%
  \Doi{10.1088/0004-637X/691/1/569}{\bibinfo {journal} {Astrophys. J.}}\ }%
  \textbf{\bibinfo {volume} {691}},\ \bibinfo {pages} {569} (\bibinfo {year}
  {2009}),\ \Eprint{http://arxiv.org/abs/0805.2632}{arXiv:0805.2632
  [astro-ph]}%
  \bibAnnoteFile{NoStop}{Jeong:2008rj}%
\bibitem{Giannantonio:2009ak}%
  \BibitemOpen
  \bibfield{author}{%
  \bibinfo {author} {\bibfnamefont{T.}~\bibnamefont{Giannantonio}}\ and\
  \bibinfo {author} {\bibfnamefont{C.}~\bibnamefont{Porciani}},\ }%
  \bibfield{journal}{%
  \Doi{10.1103/PhysRevD.81.063530}{\bibinfo {journal} {Phys. Rev.}}\ }%
  \textbf{\bibinfo {volume} {D81}},\ \bibinfo {pages} {063530} (\bibinfo {year}
  {2010}),\ \Eprint{http://arxiv.org/abs/0911.0017}{arXiv:0911.0017
  [astro-ph.CO]}%
  \bibAnnoteFile{NoStop}{Giannantonio:2009ak}%
\bibitem{Sefusatti:2009qh}%
  \BibitemOpen
  \bibfield{author}{%
  \bibinfo {author} {\bibfnamefont{E.}~\bibnamefont{Sefusatti}},\ }%
  \bibfield{journal}{%
  \Doi{10.1103/PhysRevD.80.123002}{\bibinfo {journal} {Phys. Rev.}}\ }%
  \textbf{\bibinfo {volume} {D80}},\ \bibinfo {pages} {123002} (\bibinfo {year}
  {2009}),\ \Eprint{http://arxiv.org/abs/0905.0717}{arXiv:0905.0717
  [astro-ph.CO]}%
  \bibAnnoteFile{NoStop}{Sefusatti:2009qh}%
\bibitem{Weinberg:2004kf}%
  \BibitemOpen
  \bibfield{author}{%
  \bibinfo {author} {\bibfnamefont{S.}~\bibnamefont{Weinberg}},\ }%
  \bibfield{journal}{%
  \Doi{10.1103/PhysRevD.70.083522}{\bibinfo {journal} {Phys.Rev.}}\ }%
  \textbf{\bibinfo {volume} {D70}},\ \bibinfo {pages} {083522} (\bibinfo {year}
  {2004}),\
  \Eprint{http://arxiv.org/abs/astro-ph/0405397}{arXiv:astro-ph/0405397
  [astro-ph]}%
  \bibAnnoteFile{NoStop}{Weinberg:2004kf}%
\bibitem{Weinberg:2008si}%
  \BibitemOpen
  \bibfield{author}{%
  \bibinfo {author} {\bibfnamefont{S.}~\bibnamefont{Weinberg}},\ }%
  \bibfield{journal}{%
  \Doi{10.1103/PhysRevD.79.043504}{\bibinfo {journal} {Phys.Rev.}}\ }%
  \textbf{\bibinfo {volume} {D79}},\ \bibinfo {pages} {043504} (\bibinfo {year}
  {2009}),\ \Eprint{http://arxiv.org/abs/arXiv:0810.2831}{arXiv:arXiv:0810.2831
  [hep-ph]}%
  \bibAnnoteFile{NoStop}{Weinberg:2008si}%
\bibitem{Vernizzi:2006ve}%
  \BibitemOpen
  \bibfield{author}{%
  \bibinfo {author} {\bibfnamefont{F.}~\bibnamefont{Vernizzi}}\ and\ \bibinfo
  {author} {\bibfnamefont{D.}~\bibnamefont{Wands}},\ }%
  \bibfield{journal}{%
  \Doi{10.1088/1475-7516/2006/05/019}{\bibinfo {journal} {JCAP}}\ }%
  \textbf{\bibinfo {volume} {0605}},\ \bibinfo {pages} {019} (\bibinfo {year}
  {2006}),\
  \Eprint{http://arxiv.org/abs/astro-ph/0603799}{arXiv:astro-ph/0603799
  [astro-ph]}%
  \bibAnnoteFile{NoStop}{Vernizzi:2006ve}%
\bibitem{GarciaBellido:1995qq}%
  \BibitemOpen
  \bibfield{author}{%
  \bibinfo {author} {\bibfnamefont{J.}~\bibnamefont{Garcia-Bellido}}\ and\
  \bibinfo {author} {\bibfnamefont{D.}~\bibnamefont{Wands}},\ }%
  \bibfield{journal}{%
  \Doi{10.1103/PhysRevD.53.5437}{\bibinfo {journal} {Phys.Rev.}}\ }%
  \textbf{\bibinfo {volume} {D53}},\ \bibinfo {pages} {5437} (\bibinfo {year}
  {1996}),\
  \Eprint{http://arxiv.org/abs/astro-ph/9511029}{arXiv:astro-ph/9511029
  [astro-ph]}%
  \bibAnnoteFile{NoStop}{GarciaBellido:1995qq}%
\bibitem{Choi:2007su}%
  \BibitemOpen
  \bibfield{author}{%
  \bibinfo {author} {\bibfnamefont{K.-Y.}\ \bibnamefont{Choi}}, \bibinfo
  {author} {\bibfnamefont{L.~M.}\ \bibnamefont{Hall}},\ and\ \bibinfo {author}
  {\bibfnamefont{C.}~\bibnamefont{van~de Bruck}},\ }%
  \bibfield{journal}{%
  \Doi{10.1088/1475-7516/2007/02/029}{\bibinfo {journal} {JCAP}}\ }%
  \textbf{\bibinfo {volume} {0702}},\ \bibinfo {pages} {029} (\bibinfo {year}
  {2007}),\ \bibinfo {note} {* Brief entry *},\
  \Eprint{http://arxiv.org/abs/astro-ph/0701247}{arXiv:astro-ph/0701247
  [astro-ph]}%
  \bibAnnoteFile{NoStop}{Choi:2007su}%
\bibitem{Starobinsky:1986fxa}%
  \BibitemOpen
  \bibfield{author}{%
  \bibinfo {author} {\bibfnamefont{A.~A.}\ \bibnamefont{Starobinsky}},\ }%
  \bibfield{journal}{%
  \bibinfo {journal} {JETP Lett.}\ }%
  \textbf{\bibinfo {volume} {42}},\ \bibinfo {pages} {152} (\bibinfo {year}
  {1985})%
  \bibAnnoteFile{NoStop}{Starobinsky:1986fxa}%
\bibitem{Sasaki:1995aw}%
  \BibitemOpen
  \bibfield{author}{%
  \bibinfo {author} {\bibfnamefont{M.}~\bibnamefont{Sasaki}}\ and\ \bibinfo
  {author} {\bibfnamefont{E.~D.}\ \bibnamefont{Stewart}},\ }%
  \bibfield{journal}{%
  \Doi{10.1143/PTP.95.71}{\bibinfo {journal} {Prog. Theor. Phys.}}\ }%
  \textbf{\bibinfo {volume} {95}},\ \bibinfo {pages} {71} (\bibinfo {year}
  {1996}),\
  \Eprint{http://arxiv.org/abs/astro-ph/9507001}{arXiv:astro-ph/9507001}%
  \bibAnnoteFile{NoStop}{Sasaki:1995aw}%
\bibitem{Lyth:2004gb}%
  \BibitemOpen
  \bibfield{author}{%
  \bibinfo {author} {\bibfnamefont{D.~H.}\ \bibnamefont{Lyth}}, \bibinfo
  {author} {\bibfnamefont{K.~A.}\ \bibnamefont{Malik}},\ and\ \bibinfo {author}
  {\bibfnamefont{M.}~\bibnamefont{Sasaki}},\ }%
  \bibfield{journal}{%
  \Doi{10.1088/1475-7516/2005/05/004}{\bibinfo {journal} {JCAP}}\ }%
  \textbf{\bibinfo {volume} {0505}},\ \bibinfo {pages} {004} (\bibinfo {year}
  {2005}),\
  \Eprint{http://arxiv.org/abs/astro-ph/0411220}{arXiv:astro-ph/0411220}%
  \bibAnnoteFile{NoStop}{Lyth:2004gb}%
\bibitem{Lyth:2005fi}%
  \BibitemOpen
  \bibfield{author}{%
  \bibinfo {author} {\bibfnamefont{D.~H.}\ \bibnamefont{Lyth}}\ and\ \bibinfo
  {author} {\bibfnamefont{Y.}~\bibnamefont{Rodriguez}},\ }%
  \bibfield{journal}{%
  \Doi{10.1103/PhysRevLett.95.121302}{\bibinfo {journal} {Phys. Rev. Lett.}}\
  }%
  \textbf{\bibinfo {volume} {95}},\ \bibinfo {pages} {121302} (\bibinfo {year}
  {2005}),\
  \Eprint{http://arxiv.org/abs/astro-ph/0504045}{arXiv:astro-ph/0504045}%
  \bibAnnoteFile{NoStop}{Lyth:2005fi}%
\bibitem{Bardeen:1980kt}%
  \BibitemOpen
  \bibfield{author}{%
  \bibinfo {author} {\bibfnamefont{J.~M.}\ \bibnamefont{Bardeen}},\ }%
  \bibfield{journal}{%
  \Doi{10.1103/PhysRevD.22.1882}{\bibinfo {journal} {Phys.Rev.}}\ }%
  \textbf{\bibinfo {volume} {D22}},\ \bibinfo {pages} {1882} (\bibinfo {year}
  {1980})%
  \bibAnnoteFile{NoStop}{Bardeen:1980kt}%
\bibitem{Lyth:1998xn}%
  \BibitemOpen
  \bibfield{author}{%
  \bibinfo {author} {\bibfnamefont{D.~H.}\ \bibnamefont{Lyth}}\ and\ \bibinfo
  {author} {\bibfnamefont{A.}~\bibnamefont{Riotto}},\ }%
  \bibfield{journal}{%
  \Doi{10.1016/S0370-1573(98)00128-8}{\bibinfo {journal} {Phys.Rept.}}\ }%
  \textbf{\bibinfo {volume} {314}},\ \bibinfo {pages} {1} (\bibinfo {year}
  {1999}),\ \Eprint{http://arxiv.org/abs/hep-ph/9807278}{arXiv:hep-ph/9807278
  [hep-ph]}%
  \bibAnnoteFile{NoStop}{Lyth:1998xn}%
\bibitem{Komatsu:2001rj}%
  \BibitemOpen
  \bibfield{author}{%
  \bibinfo {author} {\bibfnamefont{E.}~\bibnamefont{Komatsu}}\ and\ \bibinfo
  {author} {\bibfnamefont{D.~N.}\ \bibnamefont{Spergel}},\ }%
  \bibfield{journal}{%
  \Doi{10.1103/PhysRevD.63.063002}{\bibinfo {journal} {Phys. Rev.}}\ }%
  \textbf{\bibinfo {volume} {D63}},\ \bibinfo {pages} {063002} (\bibinfo {year}
  {2001}),\
  \Eprint{http://arxiv.org/abs/astro-ph/0005036}{arXiv:astro-ph/0005036}%
  \bibAnnoteFile{NoStop}{Komatsu:2001rj}%
\bibitem{Zaballa:2006pv}%
  \BibitemOpen
  \bibfield{author}{%
  \bibinfo {author} {\bibfnamefont{I.}~\bibnamefont{Zaballa}}, \bibinfo
  {author} {\bibfnamefont{Y.}~\bibnamefont{Rodriguez}},\ and\ \bibinfo {author}
  {\bibfnamefont{D.~H.}\ \bibnamefont{Lyth}},\ }%
  \bibfield{journal}{%
  \Doi{10.1088/1475-7516/2006/06/013}{\bibinfo {journal} {JCAP}}\ }%
  \textbf{\bibinfo {volume} {0606}},\ \bibinfo {pages} {013} (\bibinfo {year}
  {2006}),\
  \Eprint{http://arxiv.org/abs/astro-ph/0603534}{arXiv:astro-ph/0603534
  [astro-ph]}%
  \bibAnnoteFile{NoStop}{Zaballa:2006pv}%
\bibitem{Suyama:2007bg}%
  \BibitemOpen
  \bibfield{author}{%
  \bibinfo {author} {\bibfnamefont{T.}~\bibnamefont{Suyama}}\ and\ \bibinfo
  {author} {\bibfnamefont{M.}~\bibnamefont{Yamaguchi}},\ }%
  \bibfield{journal}{%
  \Doi{10.1103/PhysRevD.77.023505}{\bibinfo {journal} {Phys. Rev.}}\ }%
  \textbf{\bibinfo {volume} {D77}},\ \bibinfo {pages} {023505} (\bibinfo {year}
  {2008}),\ \Eprint{http://arxiv.org/abs/0709.2545}{arXiv:0709.2545
  [astro-ph]}%
  \bibAnnoteFile{NoStop}{Suyama:2007bg}%
\bibitem{Sugiyama:2011jt}%
  \BibitemOpen
  \bibfield{author}{%
  \bibinfo {author} {\bibfnamefont{N.~S.}\ \bibnamefont{Sugiyama}}, \bibinfo
  {author} {\bibfnamefont{E.}~\bibnamefont{Komatsu}},\ and\ \bibinfo {author}
  {\bibfnamefont{T.}~\bibnamefont{Futamase}}}%
   (\bibinfo {year} {2011}),\
  \Eprint{http://arxiv.org/abs/1101.3636}{arXiv:1101.3636 [gr-qc]}%
  \bibAnnoteFile{NoStop}{Sugiyama:2011jt}%
\bibitem{Gordon:2000hv}%
  \BibitemOpen
  \bibfield{author}{%
  \bibinfo {author} {\bibfnamefont{C.}~\bibnamefont{Gordon}}, \bibinfo {author}
  {\bibfnamefont{D.}~\bibnamefont{Wands}}, \bibinfo {author}
  {\bibfnamefont{B.~A.}\ \bibnamefont{Bassett}},\ and\ \bibinfo {author}
  {\bibfnamefont{R.}~\bibnamefont{Maartens}},\ }%
  \bibfield{journal}{%
  \Doi{10.1103/PhysRevD.63.023506}{\bibinfo {journal} {Phys.Rev.}}\ }%
  \textbf{\bibinfo {volume} {D63}},\ \bibinfo {pages} {023506} (\bibinfo {year}
  {2001}),\
  \Eprint{http://arxiv.org/abs/astro-ph/0009131}{arXiv:astro-ph/0009131
  [astro-ph]}%
  \bibAnnoteFile{NoStop}{Gordon:2000hv}%
\bibitem{Wanatanabe:2009}%
  \BibitemOpen
  \bibfield{author}{%
  \bibinfo {author} {\bibfnamefont{Y.}~\bibnamefont{Watanabe}},\ }%
  \emph{\bibinfo {title} {{Toward Understanding of the Complete Thermal History
  of the Universe by Gravitation}}},\ Ph.D. thesis,\ \bibinfo {school} {The
  University of Texas at Austin} (\bibinfo {year} {2009})%
  \bibAnnoteFile{NoStop}{Wanatanabe:2009}%
\bibitem{Riotto:2002yw}%
  \BibitemOpen
  \bibfield{author}{%
  \bibinfo {author} {\bibfnamefont{A.}~\bibnamefont{Riotto}},\ \bibinfo {pages}
  {317}}%
   (\bibinfo {year} {2002}),\
  \Eprint{http://arxiv.org/abs/hep-ph/0210162}{arXiv:hep-ph/0210162 [hep-ph]}%
  \bibAnnoteFile{NoStop}{Riotto:2002yw}%
\bibitem{Byrnes:2008wi}%
  \BibitemOpen
  \bibfield{author}{%
  \bibinfo {author} {\bibfnamefont{C.~T.}\ \bibnamefont{Byrnes}}, \bibinfo
  {author} {\bibfnamefont{K.-Y.}\ \bibnamefont{Choi}},\ and\ \bibinfo {author}
  {\bibfnamefont{L.~M.~H.}\ \bibnamefont{Hall}},\ }%
  \bibfield{journal}{%
  \Doi{10.1088/1475-7516/2008/10/008}{\bibinfo {journal} {JCAP}}\ }%
  \textbf{\bibinfo {volume} {0810}},\ \bibinfo {pages} {008} (\bibinfo {year}
  {2008}),\ \Eprint{http://arxiv.org/abs/0807.1101}{arXiv:0807.1101
  [astro-ph]}%
  \bibAnnoteFile{NoStop}{Byrnes:2008wi}%
\bibitem{Kim:2010ud}%
  \BibitemOpen
  \bibfield{author}{%
  \bibinfo {author} {\bibfnamefont{S.~A.}\ \bibnamefont{Kim}}, \bibinfo
  {author} {\bibfnamefont{A.~R.}\ \bibnamefont{Liddle}},\ and\ \bibinfo
  {author} {\bibfnamefont{D.}~\bibnamefont{Seery}},\ }%
  \bibfield{journal}{%
  \Doi{10.1103/PhysRevLett.105.181302}{\bibinfo {journal} {Phys.Rev.Lett.}}\ }%
  \textbf{\bibinfo {volume} {105}},\ \bibinfo {pages} {181302} (\bibinfo {year}
  {2010}),\ \Eprint{http://arxiv.org/abs/1005.4410}{arXiv:1005.4410
  [astro-ph.CO]}%
  \bibAnnoteFile{NoStop}{Kim:2010ud}%
\bibitem{Fry:1983cj}%
  \BibitemOpen
  \bibfield{author}{%
  \bibinfo {author} {\bibfnamefont{J.~N.}\ \bibnamefont{Fry}},\ }%
  \bibfield{journal}{%
  \Doi{10.1086/161913}{\bibinfo {journal} {Astrophys.J.}}\ }%
  \textbf{\bibinfo {volume} {279}},\ \bibinfo {pages} {499} (\bibinfo {year}
  {1984})%
  \bibAnnoteFile{NoStop}{Fry:1983cj}%
\bibitem{riordan2002introduction}%
  \BibitemOpen
  \bibfield{author}{%
  \bibinfo {author} {\bibfnamefont{J.}~\bibnamefont{Riordan}},\ }%
  \emph{\bibinfo {title} {Introduction to Combinatorial Analysis}}\ (\bibinfo
  {publisher} {Dover Publications},\ \bibinfo {year} {2002})%
  \bibAnnoteFile{NoStop}{riordan2002introduction}%
\bibitem{Cogollo:2008bi}%
  \BibitemOpen
  \bibfield{author}{%
  \bibinfo {author} {\bibfnamefont{H.~R.}\ \bibnamefont{Cogollo}}, \bibinfo
  {author} {\bibfnamefont{Y.}~\bibnamefont{Rodriguez}},\ and\ \bibinfo {author}
  {\bibfnamefont{C.~A.}\ \bibnamefont{Valenzuela-Toledo}},\ }%
  \bibfield{journal}{%
  \Doi{10.1088/1475-7516/2008/08/029}{\bibinfo {journal} {JCAP}}\ }%
  \textbf{\bibinfo {volume} {0808}},\ \bibinfo {pages} {029} (\bibinfo {year}
  {2008}),\ \Eprint{http://arxiv.org/abs/0806.1546}{arXiv:0806.1546
  [astro-ph]}%
  \bibAnnoteFile{NoStop}{Cogollo:2008bi}%
\bibitem{Rodriguez:2008hy}%
  \BibitemOpen
  \bibfield{author}{%
  \bibinfo {author} {\bibfnamefont{Y.}~\bibnamefont{Rodriguez}}\ and\ \bibinfo
  {author} {\bibfnamefont{C.~A.}\ \bibnamefont{Valenzuela-Toledo}},\ }%
  \bibfield{journal}{%
  \Doi{10.1103/PhysRevD.81.023531}{\bibinfo {journal} {Phys.Rev.}}\ }%
  \textbf{\bibinfo {volume} {D81}},\ \bibinfo {pages} {023531} (\bibinfo {year}
  {2010}),\ \Eprint{http://arxiv.org/abs/0811.4092}{arXiv:0811.4092
  [astro-ph]}%
  \bibAnnoteFile{NoStop}{Rodriguez:2008hy}%
\bibitem{Byrnes:2007tm}%
  \BibitemOpen
  \bibfield{author}{%
  \bibinfo {author} {\bibfnamefont{C.~T.}\ \bibnamefont{Byrnes}}, \bibinfo
  {author} {\bibfnamefont{K.}~\bibnamefont{Koyama}}, \bibinfo {author}
  {\bibfnamefont{M.}~\bibnamefont{Sasaki}},\ and\ \bibinfo {author}
  {\bibfnamefont{D.}~\bibnamefont{Wands}},\ }%
  \bibfield{journal}{%
  \Doi{10.1088/1475-7516/2007/11/027}{\bibinfo {journal} {JCAP}}\ }%
  \textbf{\bibinfo {volume} {0711}},\ \bibinfo {pages} {027} (\bibinfo {year}
  {2007}),\ \Eprint{http://arxiv.org/abs/0705.4096}{arXiv:0705.4096 [hep-th]}%
  \bibAnnoteFile{NoStop}{Byrnes:2007tm}%
\end{thebibliography}%

\end{document}